\documentclass[fleqn,usenatbib]{mn2e}
\usepackage{graphicx,amsmath,amssymb}
\usepackage[usenames,dvipsnames]{xcolor}
\pdfoutput=1 % For arXiv
\usepackage{hyperref}
\usepackage{natbib}
\usepackage{times}
\newcommand{\degrees}{\ensuremath{^{\circ}}}
\newcommand{\NII}{[\ion{N}{ii}]}
\newcommand{\SII}{[\ion{S}{ii}]}
\newcommand{\Halpha}{H~\ensuremath{\alpha}}
\newcommand{\Hbeta}{H~\ensuremath{\beta}}
\newcommand{\Lyalpha}{Ly~\ensuremath{\alpha}}
\newcommand{\fbfn}{\ensuremath{f_\mathrm{b}}/\ensuremath{f_\mathrm{n}}}

\newcommand{\kms}{km~s$^{-1}$}
\newcommand{\vmax}{\ensuremath{\Delta v_\mathrm{max}}}
\newcommand{\vmaxone}{\ensuremath{\Delta v_\mathrm{max,1}}}
\newcommand{\vmaxtwo}{\ensuremath{\Delta v_\mathrm{max,2}}}

\newcommand{\OonH}{\ensuremath{12+\log(\mathrm{O/H})}}

\newcommand{\ion}[2]{#1~\textsc{#2}}

\title[Supernova-Driven Outflows in NGC 7552]{Supernova-Driven Outflows in NGC 7552: A Comparison of \Halpha\ and UV Tracers}
\author[Corey M.\ Wood et al.]
    {Corey M.\ Wood,$^1$\thanks{E-mail: \href{mailto:wood@astro.wisc.edu}{wood@astro.wisc.edu}}
     Christy A.\ Tremonti,$^1$
     Daniela Calzetti,$^2$
     Claus Leitherer,$^3$
     \newauthor
     John Chisholm,$^1$
     and John S.\ Gallagher \textsc{iii}$^1$
     \\
$^1$Department of Astronomy, University of Wisconsin--Madison, 475 N.\ Charter St., Madison, WI 53706, USA\\
$^2$Department of Astronomy, University of Massachusetts, Amherst, MA 01003, USA\\
$^3$Space Telescope Science Institute, 3700 San Martin Drive, Baltimore, MD 21218, USA
}

\begin{document}
\date{\today}
\pagerange{\pageref{firstpage}--\pageref{lastpage}} \pubyear{2015}
\maketitle
\label{firstpage}

\begin{abstract}
We investigate the supernova-driven galactic wind of the barred spiral galaxy NGC~7552, using both ground-based optical nebular emission lines and far-ultraviolet absorption lines measured with the \emph{Hubble Space Telescope} Cosmic Origins Spectrograph. We detect broad ($\sim300$~km~s$^{-1}$) blueshifted ($-40$~km~s$^{-1}$) optical emission lines associated with the galaxy's kpc-scale star-forming ring. The broad line kinematics and diagnostic line ratios suggest that the H~$\alpha$ emission comes from clouds of high density gas entrained in a turbulent outflow. We compare the H~$\alpha$ emission line profile to the UV absorption line profile measured along a coincident sight line and find significant differences. The maximum blueshift of the H~$\alpha$-emitting gas is $\sim290$~km~s$^{-1}$, whereas the UV line profile extends to blueshifts upwards of 1000~km~s$^{-1}$. The mass outflow rate estimated from the UV is roughly nine times greater than that estimated from H~$\alpha$. We argue that the H~$\alpha$ emission traces a cluster-scale outflow of dense, low velocity gas at the base of the large-scale wind. We suggest that UV absorption line measurements are therefore more reliable tracers of warm gas in starburst-driven outflows.
\end{abstract}


\section{Introduction}
\label{sec:intro}
Feedback from massive stellar winds and supernova explosions has long been identified as a mechanism capable of injecting large amounts of energy into the local interstellar medium (ISM) of a galaxy, resulting in both the heating and potential removal of the gas in the form of a wind \citep{Larson74}.
Supernovae-driven winds have myriad profound effects on galaxy evolution, influencing the shape of the galaxy luminosity function \citep{Benson03}, the mass-metallicity relation \citep{Finlator08}, and the structure of galactic disks \citep{Scannapieco08}.
Such winds have been found to be ``ubiquitous'' in star-forming galaxies where star formation surface densities exceed $\Sigma_{SFR} \approx 0.1$~M$_\odot$~yr$^{-1}$~kpc$^{-2}$ \citep{Heckman02}.
Under these conditions, rapid star formation results in a large injection of mechanical energy into the local ISM of the galaxy by OB stars, Wolf-Rayet stars, and supernovae.

Although galactic winds are commonly observed both locally and, increasingly, in the high-redshift universe \citep[e.g.,][]{Veilleux05, Steidel10}, it has been difficult to determine how much mass these winds remove from their host galaxies.
The extent to which galactic winds affect on-going star formation depends highly on the ability of such winds to remove gas from their hosts, commonly parameterized as the mass loading factor $\eta = \dot{M}_\mathrm{out} / \dot{M}_\mathrm{*}$, the rate of mass loss due to an outflow as a fraction of the global star formation rate.
Measured mass loading factors are typically around $\eta \sim 1$, but many of these measurements come with high uncertainties.
\citet{Rupke05} measure mass loading factors of $\eta \approx$ 0.01 -- 1 in $\sim45$ starburst-dominated galaxies at $z < 0.5$.
These measurements suffer from order-of-magnitude uncertainties due to large ionization corrections for \ion{Na}{i} absorption lines.
\citet{Bouche12} measure $\eta \sim 2$ with uncertainties of only a factor of 2 for five galaxies studied via background quasar absorption, but warn that it may be incorrect to compare outflows measured at large impact parameter to the current level of star formation.
At higher redshift, \citet{Pettini02} estimate $\eta \sim 1$ in a single Lyman break galaxy at $z = 2.73$.
\citet{Newman12b} measure $\eta \sim 2$ for high-$\Sigma_{SFR}$ systems at $z \sim 2$, but these measurements suffer from quoted uncertainties of at least a factor of 3.
The uncertainties are possibly much larger due to uncertainties in the electron density measurement because of low-S/N in the \SII\ lines.
More robust measurements of outflow masses and velocities will provide better constraints on mass loading factors.

Mass loss measurements have typically been easier to perform in absorption-line studies, where the column density of the absorbing gas is readily recovered from unsaturated lines.
Absorption studies are well-suited to face-on systems where the galaxy serves as a background source.
The most common absorption probe in the easily-accessible visible wavelength region is the \ion{Na}{i~D} $\lambda\lambda$5889,5896 doublet, which traces cool, neutral gas \citep[e.g.,][]{Chen10}.
But the utility of this line is limited due to its relatively low ionization potential ($\sim5$~eV), leaving it highly susceptible to ionization from ultraviolet (UV) sources and making it most useful in studies of dense, dusty environments \citep{Martin05, Rupke05}.
Even when neutral \ion{Na}{i} is present, ionization corrections are highly uncertain, leading to large uncertainties in total gas mass calculations.
More robust absorption measurements can be made in the UV using lines like \ion{Mg}{ii} $\lambda\lambda$2796, 2804 or \ion{Fe}{ii} $\lambda\lambda$2586, 2600.
These lines have been used to study galactic winds at moderate redshifts \citep[$z \sim$ 0.5 -- 1.5; e.g.,][]{Tremonti07, Weiner09} where these lines fall in the visible regime, but in the local universe these lines require observations from space telescopes, limiting the availability of such data.
UV observations carry the additional requirement that observed targets must be UV-bright.

Emission-line studies have typically been limited to edge-on systems where the geometry naturally separates extra-planar emission from that of the disk.
Wind energetics and mass estimates are substantially more difficult to measure in edge-on systems, however, due to the outflow velocities being projected into the plane of the sky.
Emission line studies in face-on systems are unable to spatially separate nebular lines in disk \ion{H}{ii} regions from extra-planar outflow gas.
Wind emission will have distinct kinematic features at modest spectral and spatial resolution, allowing for the separation of the two components in velocity space.
Spatially resolving the galactic rotation curve minimizes velocity smearing between spectral apertures, allowing for more accurate recovery of multiple velocity components within a single line profile.
Galaxies at low inclination are free of strong projection effects, allowing for more accurate spatial mapping of an outflow.

Recent estimates of ionized mass outflow have been made in high redshift galaxies ($z\sim2$) using rest-frame \Halpha\ emission lines \citep{Genzel11, Newman12a}.
These methods rely on decomposing the \Halpha\ emission line into two velocity components, a narrow component and a broad component.
This broad component is interpreted as being generated by the outflowing gas, and can thus be used as a direct measurement of the outflow properties.
This method has also been used recently in the study of high-z AGN outflows \citep{Genzel14} and luminous and ultra-luminous infrared galaxies \citep[LIRG, $L_{IR} > 10^{11} L_\odot$; ULIRG, $L_{IR} > 10^{12} L_\odot$;][]{Arribas14}.
This method is promising in these high-redshift studies but requires further testing in the local universe, where spatially-resolved observations are possible at higher signal-to-noise.

These \Halpha-based methods for estimating the mass outflow rate also require testing against absorption-line methods, as absorption-line measurements are sensitive to a wider range of column densities.
Absorption-line studies therefore may still be sensitive to gas unseen in \Halpha\ observations.
A direct comparison of emission and absorption data would provide insight into the relative mass ratios of gas traced by these different observational methods.

In this paper, we analyze the galactic wind in the nearby face-on spiral galaxy NGC~7552.
We analyze visible long-slit spectra to measure the properties and mass outflow rate of the wind using nebular emission lines, and investigate the validity of the assumption that the broad velocity component traces the galactic wind.
We also calculate an independent mass outflow rate using UV absorption lines, and directly compare these two results.
We compare these values to other similar works, including studies of systems at high redshift.
In \S\ref{sec:7552} we present a general overview of the galaxy NGC~7552.
In \S\ref{sec:data} we describe our data sources and detail our analysis methods.
We examine the results of our analysis in \S\ref{sec:results} and \S\ref{sec:uv_results}, and discuss broader implications in \S\ref{sec:discussion}.
Finally, we summarize our conclusions in \S\ref{sec:conclusion}.
For this work we adopt $H_0 = 73$~km~s$^{-1}$~Mpc$^{-1}$, $\Omega_m = 0.27$, and $\Omega_\Lambda = 0.73$.

\section{NGC 7552}
\label{sec:7552}
NGC~7552 is a nearby barred spiral galaxy in the Grus Quartet, a group of four large spiral galaxies in the southern sky.
It lies at an estimated distance of $22.5 \pm 1.6$~Mpc based on a three-attractor flow model \citep[$1\arcsec = 0.109$~kpc;][]{Mould00}, with a heliocentric radial velocity of 1586~\kms\ \citep{Pan13}.
The galaxy is seen largely face-on at an inclination of only 23 -- 28\degrees\ \citep{Feinstein90, Pan13}.
Although the morphology of NGC~7552 does not appear highly disturbed, \ion{H}{i} observations of the galaxy group show numerous tidal features indicative of interactions between the group members \citep{Freeland09}.
NGC~7552 has a galactic stellar mass of $\log M_* = 10.52$, measured by the \emph{Spitzer} Survey of the Stellar Structure in Galaxies (S$^4$G) via a combination of the 3.6 $\mu$m and 4.5 $\mu$m imaging bands on the \emph{Spitzer Space Telescope} \citep{Sheth10}.
\citet{Moustakas10} report a metallicity of $\OonH = 9.16\pm0.01$ using the ``strong line'' photoionization scaling, suggesting the galaxy is substantially super-solar ($Z \approx 3 Z_\odot$).
They also report measurements tied to the empirical metallicity scale that are 0.81 dex lower, decidedly sub-solar.
However, we have found in our analysis that a twice-solar-metallicity stellar continuum model is a good match to our data (see \S\S\ref{subsec:vis}, \ref{subsec:uv}).
Despite the ambiguity in measurements of $\OonH$, we believe that the galaxy is of at least solar metallicity.

NGC~7552 has been classified as a ``low-ionization nuclear emission-line region'' (LINER) based on its [\ion{O}{i}] emission \citep{Durret88}, raising questions of whether it hosts an active galactic nucleus (AGN).
\citet{Grier11} measure a nuclear X-ray luminosity of $L_{0.3-8~\mathrm{keV}} = 1.73\times10^{40}$~erg~s$^{-1}$, corrected for our assumed distance.
This is well-below the X-ray luminosity of nearby low-luminosity AGN \citep[e.g.,][]{Asmus11} and thus we infer that if an AGN is present, it is a minor contributor to the galaxy's bolometric luminosity.

The galaxy just barely qualifies as a LIRG with a total infrared luminosity of $L_{IR} = 1.2\times10^{11} L_{\odot}$ \citep[again corrected for our assumed distance]{Sanders03}.
Using the calibration of \citet{Calzetti13}, which assumes a Kroupa IMF \citep{Kroupa01}, this luminosity corresponds to a star formation rate of $\dot{M_*} = 13$~M$_\odot$~yr$^{-1}$.
Another approach is to use both the total infrared luminosity and the far-ultraviolet luminosity in order to account for both obscured and unobscured star formation.
Using measurements from the Galaxy Evolution Explorer \citep[\emph{GALEX};][]{GALEX}, we adopt a far-UV luminosity of $L_{FUV} = 2.38\times10^{9} L_\odot$ \citep{GilDePaz07}.
Combined with $L_{IR}$ from above, and using the IR+FUV calibrations of \citet{Hao11}, this measurement yields $\dot{M_*} = 9.9$~M$_\odot$~yr$^{-1}$.
Various other indicators yield global SFRs from 15 -- 47~M$_\odot$~yr$^{-1}$ \citep[and further references therein]{Pan13}.

One of the defining morphological features of NGC~7552 is its circum-nuclear starburst ring \citep[e.g.,][]{Feinstein90, Forbes94}.
This ring is approximately 0.5~kpc in radius and consists of numerous ``knots'' of heavy, embedded star formation.
The SFR within this ring alone is as high as 10--15~M$_{\odot}$~yr$^{-1}$, with a total infrared luminosity of $5.5\times10^{10} L_\odot$~\citep[and further references therein]{Pan13}.
Using near- and mid-infrared observations, \citet{Brandl12} identify nine prominent star clusters within the ring, with stellar ages ranging between 5.6~Myr and 6.3~Myr.

\section{Data}
\label{sec:data}
\begin{figure}
    \centering
    \includegraphics[width=\columnwidth]{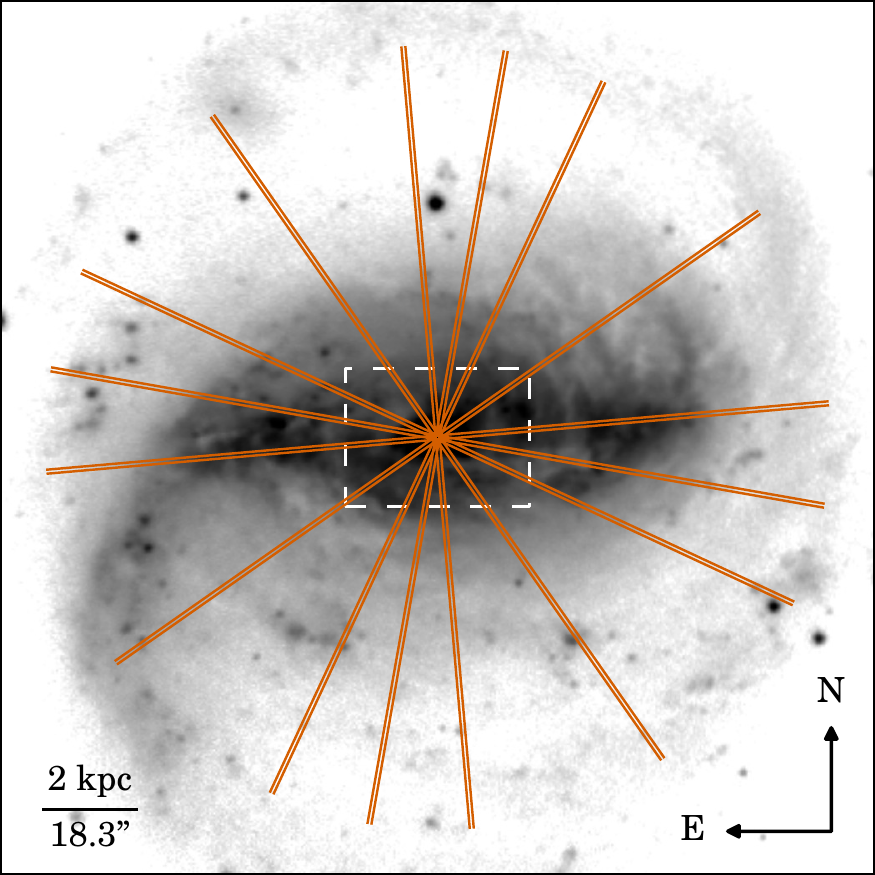}
    \caption{
        An \emph{R}-band image of NGC~7552 from the SINGS project overlaid with our eight 0\farcs8-wide slit positions in orange.
        The white dashed rectangle denotes the 4~kpc $\times$ 3~kpc region detailed in Figure~\ref{fig:Ha_map}. \label{fig:slits}}
\end{figure}

\subsection{Visible}
\label{subsec:vis}
Our main data set consists of 16 long-slit spectra obtained at eight different position angles, each centered on the nucleus, using a 0\farcs8-wide slit.
The eight position angles range from 35\degrees--185\degrees, providing coarse two-dimensional spatial information across the face of the galaxy (see Figure~\ref{fig:slits}).
Our data were obtained using the 100-inch Ir\'en\'ee du Pont telescope at Las Campanas Observatory using the Modular Spectrograph instrument. % Can't find a cite for the instrument.
The CRAF CCD with 12~$\mu$m pixels was used, giving a native plate scale of 0\farcs292~pix$^{-1}$ before applying $2 \times 2$ on-chip binning.
The observation dates span 1996 August 15--18.
Our observed spectral bandpass is 6400--6900~\AA, covering the rest-frame \NII/\Halpha\ complex and the \SII~$\lambda\lambda$6716,6731 doublet.
Sky line measurements indicate a velocity resolution of approximately 40~\kms\ full-width-at-half-maximum (FWHM).

The data were reduced using standard \textsc{iraf} long-slit spectral reduction techniques.
The data do not include separate sky pointings, so we created sky apertures for each observation by identifying regions that were free from nebular emission under visual inspection.
The median of the rectified sky spectra was subtracted from the data.
We note that in general there are no strong sky lines in our bandpass.
We then removed these sky apertures from our subsequent analysis.
We do not apply a correction for telluric absorption, as the prominent telluric features do not overlap our lines of interest.

\begin{figure}
    \centering
    \includegraphics[width=\columnwidth]{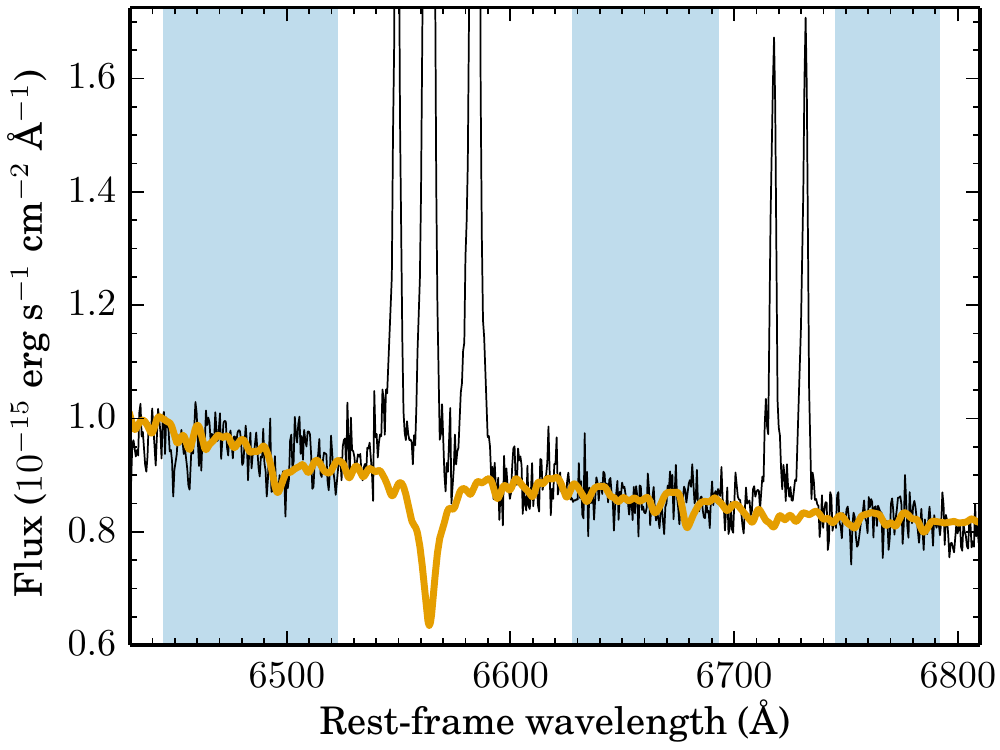}
    \caption{
        An example of the stellar continuum fit.
        A representative spectrum is shown in black and our stellar continuum model is shown in orange.
        The three blue shaded regions denote the spectral windows that were used to match the model to the observed spectrum.
        \label{fig:cont}}
\end{figure}

In order to remove the stellar continuum from our observations, we subtracted a model starburst continuum from each aperture.
We find that a twice-solar-metallicity, 100-Myr constant-SFR model from \citet{GonzalezDelgado05} adequately matches the continuum over our target spectral range.
This model accounts for deep stellar \Halpha\ absorption, providing an important correction for lines with low equivalent widths (EW $\lesssim 10$~\AA).
The model has a wavelength spacing of 0.3~\AA\ ($\sim 13$~\kms), and we then convolve the model by our instrumental resolution and approximate narrow-component line width ($\sim 100$~\kms; see \S\ref{sec:results}) to account for stellar velocity dispersion.
To match the model continuum to our data, we first used the median values within three different spectral windows to create second-order polynomial representations of both the model and the data.
We then scaled the model to the data using the ratio of the two polynomials as a pixel-to-pixel scaling factor.
This scaled model was then subtracted from the data to remove the continuum flux.
A representative example fit is shown in Figure~\ref{fig:cont}.

\begin{figure*}
    \centering
    \includegraphics[width=\textwidth]{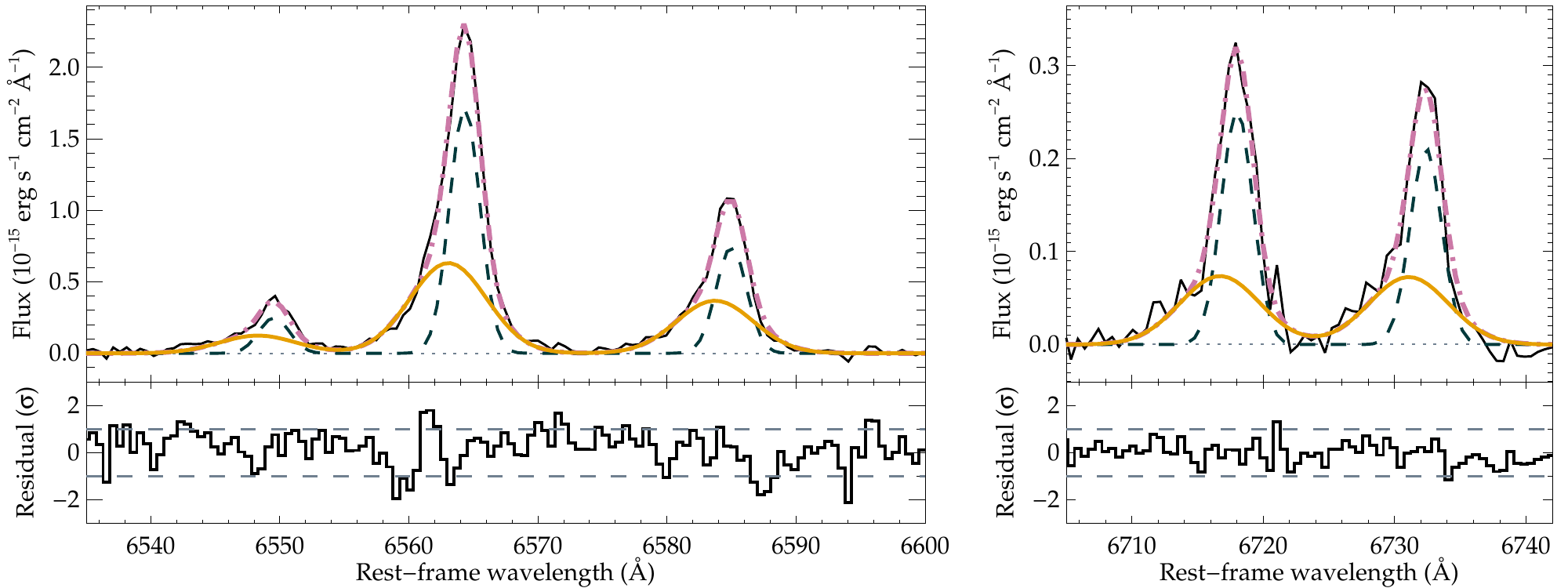}
    \caption{Line profile fits for a representative spectral aperture of our data.
             The observed spectrum is in black, the narrow and broad components are shown in dashed-blue and orange, respectively, while the purple dot-dashed line shows the total fit, or the sum of the two components.
             The dashed gray lines in the residual plots denote the $\pm1\sigma$ levels.
             \emph{Left}: The \NII/\Halpha\ complex.
             \emph{Right}: The \SII\ 6716/6731 doublet.
             \label{fig:fit}
             }
\end{figure*}

Prior to fitting the emission lines, we adaptively binned the spatial apertures to create apertures with S/N(\Halpha)~$>$~3.
This only proved necessary outside the inner 1~kpc where the galaxy's rotation curve has become fairly flat, greatly reducing the amount of velocity smearing introduced by binning these apertures.
We performed line fitting using a custom wrapper to the \textsc{mpfit} package \citep{Markwardt09}, a non-linear least squares fitting code written in the Interactive Data Language (IDL).
This custom wrapper simultaneously fits Gaussians to the five nebular emission lines within our spectral window: \NII\ $\lambda\lambda$6548,6583, \Halpha, and \SII\ $\lambda\lambda$6716,6731.
We ultimately fit the data in two passes.
We initially fit our data with a single Gaussian component for each line.
A single line component did not produce a good fit to the data, motivating the addition of a second Gaussian component to the fit.
For the second pass we simultaneously fit two Gaussian components to each emission line.
We used the results of the single-component fit as a ``first guess,'' providing initial parameters for approximate line centers, widths, and fluxes.
For our fitting we restrict the line widths of the narrow and broad components (Gaussian $\sigma < 70$~\kms\ and $71 < \sigma < 300$~\kms, respectively).
We experimented with unrestricted line widths in our fitting and found that this only has a $\sim4\%$ effect on our calculations in \S\ref{sec:results}; as such, we retain the line width restrictions to keep the narrow component and broad component easily-separated in our analysis.

For each velocity component, the flux of each spectral line is allowed to vary, while we assume each spectral line shares the same velocity center and line width.
We do, however, fix the \NII\ flux ratio at \emph{f}$_{6583}$/\emph{f}$_{6548} = 3$ in accordance with atomic physics \citep{Osterbrock}.
Line widths are also corrected for instrumental resolution ($\sigma_{inst} = 17$~\kms).
It is possible that the various emission lines do not share the same kinematics.
However, we generally find a good two-component fit ($\chi^2_{red} \sim 1$) in a large majority of apertures, and find that our data do not support additional velocity components.
Our fitting is also as-restrictive or often less-restrictive than many similar recent studies \citep[e.g.,][]{Newman12b, Genzel14, Ho14}.

An example of the result of our line fitting is shown in Figure~\ref{fig:fit}.
Our fitting routine produces good fits in almost all measured spectral apertures.
A small number of apertures within $\sim$200~pc of the galactic nucleus show larger fit residuals.
In apertures where $\chi^2_{red} > 2$ and S/N(\Halpha) $> 10$, we experimented with adding an additional broad Gaussian component to the fit.
However, this additional component in the fit proved to be unstable and did not give consistent results.
We therefore conduct our analysis solely on the basis of a two-component fit.

Because our analysis is concerned with spatially-resolved properties of the galaxy, we must ensure that our observations are properly registered with respect to each other.
Our data files do not have WCS information recorded in the headers, so a manual alignment was required.
We have no spatial information perpendicular to the slit direction, so we only investigate the centering of the galaxy along the length of the slit and assume the slit passes through the nucleus.
We made an initial estimate of the galaxy center location in each slit by selecting the aperture with the highest continuum flux.
In order to determine any required offsets from our initial estimate, we compare the continuum levels of our spectra to radial profiles extracted at each of our slit position angles from archival \emph{R}-band imaging from the SINGS project \citep{Kennicutt03}.
This comparison showed good slit alignment for most slit positions, and only small corrections (1--3 pixels) were needed for two slit positions.
As a check of our registration, we compared all the median continuum values measured at $r = 0$ and found a standard deviation of 6 percent across all slits.
The largest deviation from the mean of the median continuum levels is 10 percent.
Some slit positions show evidence of requiring sub-pixel shifts for ideal registration, but we have chosen not to interpolate our data on the sub-pixel level as our analysis does not strongly depend on the precise 2D registration of our observations.

\begin{figure}
    \includegraphics[width=\columnwidth]{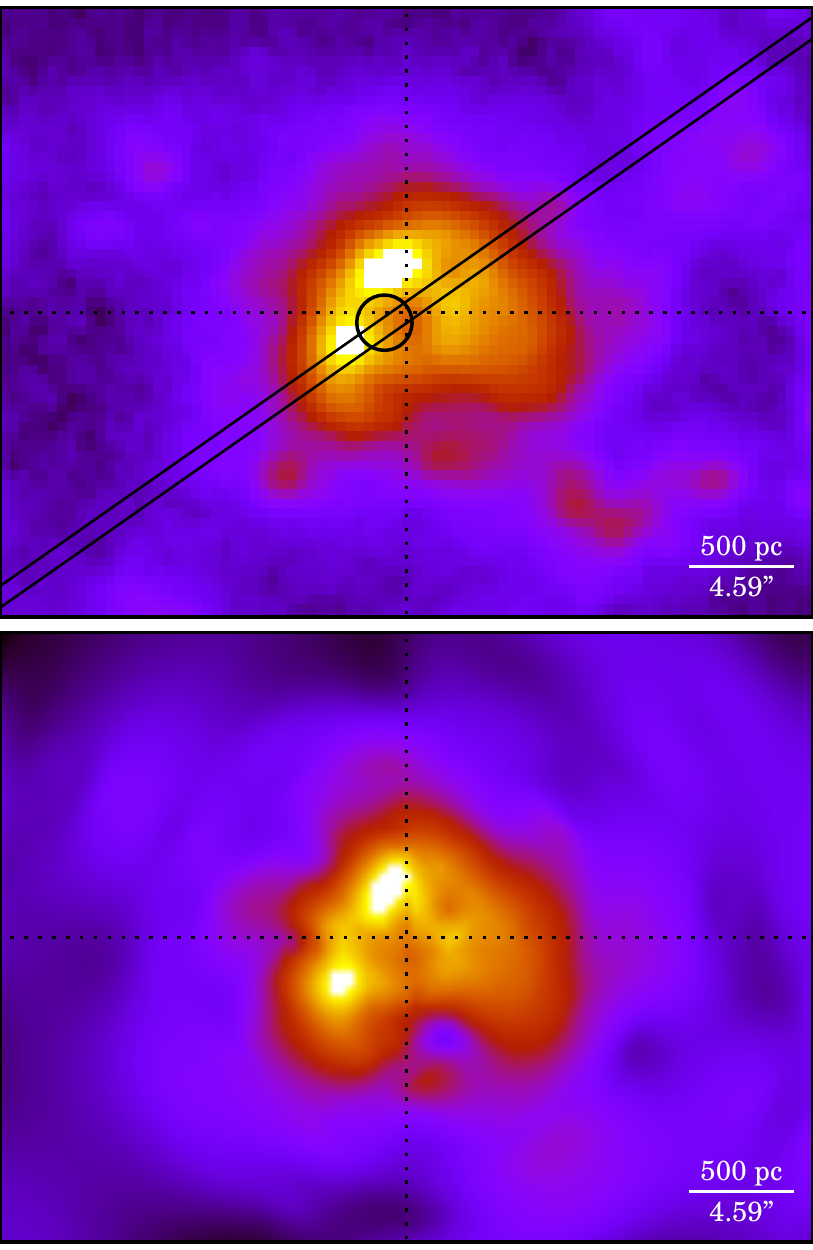}
    \caption{ \Halpha\ narrowband image of NGC~7552 from the SINGS project (top) and our 2D reconstruction of our total observed \Halpha\ flux (bottom).
        Both images use logarithmic color scaling.
        The images show the central 4~kpc $\times$ 3~kpc region of the galaxy highlighted in Fig.~\ref{fig:slits}.
        The black circle represents the COS primary science aperture (2\farcs5 diameter) at the pointing of our archival UV absorption data.
        For reference, we also show a longslit at PA $=$ 125\degrees.
        Data from this slit position are shown in Fig.~\ref{fig:voff}.
        The crosshairs mark the nucleus of the galaxy.
        \label{fig:Ha_map}
    }
\end{figure}

To further check our registration, we created an interpolated 2D map of our measured total \Halpha\ emission and compared the result to an \Halpha\ narrowband image, again from the SINGS project.
We used the \textsc{SurGe}\footnote{\url{http://surgeweb.sweb.cz}} interpolation software package for our 2D image reconstruction.
The comparison is shown in Figure~\ref{fig:Ha_map}.
The reconstruction shows good qualitative agreement with the narrowband image data.
In particular, we successfully recover the well-known star-forming ring of NGC~7552.

One goal of this work is to calculate the mass outflow rate of NGC~7552 in a manner consistent with recent studies performed at high redshift \citep[$z\sim2$; e.g.,][]{Newman12b}.
This calculation depends on the observed \Halpha\ broad-line luminosity.
To ensure accurate absolute spectrophotometric calibration, we have matched our integrated \Halpha\ counts to calibrated \Halpha\ narrow-band imagery from the SINGS project.
This image was acquired through a narrow-band \Halpha\ filter with a central wavelength of 6596~\AA\ (6561~\AA\ in the galaxy's rest-frame) and an 18~\AA\ FWHM, resulting in negligible \NII\ contamination.
The broadband image has a calibration accuracy of approximately 10 percent.

\subsection{Ultraviolet}
\label{subsec:uv}
We also use an archival ultraviolet (UV) spectrum of NGC~7552 obtained with the Cosmic Origins Spectrograph (COS) instrument \citep{Osterman11,Green12} on-board the \emph{Hubble Space Telescope} (\emph{HST}).
The spectrum was acquired as part of \emph{HST} program 12173 (PI: Leitherer), which explored the UV properties of four infrared luminous galaxies.
\citet{Leitherer13} analyzed the stellar content, Lyman-$\alpha$ profiles, and interstellar medium absorption lines of all four galaxies and estimated a mean mass outflow rate for the sample.
Here we take a closer look at the multiphase gas in NGC~7552 and focus on the comparison of outflow properties inferred from the UV and \Halpha\ lines.

The COS spectrum was acquired using the 2\farcs5-diameter Primary Science Aperture with the G130M grating.
Fortuitously, the COS pointing falls along one of our long-slit positions, allowing for a direct comparison of the same physical location in the galaxy (see Figure~\ref{fig:Ha_map}).
The data cover the wavelength region 1139.52 -- 1451.53~\AA.
The point source spectral resolution of COS/G130M is $R \approx 17,000$ (18~\kms).
Measured line widths of unresolved Milky Way foreground lines in the spectrum indicate that the achieved resolution is closer to 100~\kms FWHM.
The lower resolution is a consequence of the fact that COS is optimized for point sources and NGC~7552 fills the aperture.
Additional data exist from the Faint Object Spectrograph (FOS) and the Space Telescope Imaging Spectrograph (STIS) instruments, both also aboard \emph{HST}.
A visual inspection of this additional data shows qualitatively similar features to the COS spectrum, but at substantially lower resolution; we therefore use only the COS data in our analysis.

In order to minimize the impact of micro-channel plate detector fixed-pattern noise, NGC~7552 was observed at two central wavelength positions shifted by 18~\AA\ for a total of 5599 seconds \citep{Leitherer13}.
The individual spectra were downloaded from the MAST server and processed through the CalCOS pipeline, version 2.20.1.
The individual exposures were co-added using a procedure developed by \citet{Wakker15} to optimally align the spectra.
Strong absorption lines in individual exposures are cross-correlated to give a wavelength-dependent wavelength solution, and then the raw counts are combined and converted into flux values.
The errors are calculated from the raw gross counts, after accounting for the proper background levels (as defined in the COS handbook \citep{Holland14}).
Special attention is paid to edge effects and fixed pattern noise features, which are known to affect the COS data \citep{Danforth10}.
We correct the final spectrum for foreground Milky Way dust attenuation using $E(B - V) = 0.012$ \citep{Schlafly11} and a Milky Way dust attenuation curve from \citep{Fitzpatrick99}.

Our goal is to use the rich array of interstellar medium (ISM) absorption lines in the COS spectrum to probe gas in the outflow.
To accurately measure the ISM lines, we must first model and remove any features in the background light source: the stellar continuum.
We use the standard technique of UV spectral synthesis \citep[e.g.,][]{Tremonti01} to identify a best fit model of the stellar continuum.
We use stellar population models produced by Starburst99 \citep[SB99;][]{Leitherer99} with a Kroupa IMF, a power-law index of 1.3 for the low mass slope, an index of 2.3 for the high mass slope, a high mass cut-off at 100~M$_\odot$, and the Geneva stellar evolution tracks for non-rotating stars with high mass-loss.
We utilized the high-resolution, fully theoretical model atmospheres calculated using the WM-Basic code \citep{Leitherer10} instead of the default empirical libraries.
The theoretical libraries provide an excellent match to the spectral resolution of our data and are free of the ISM absorption lines that contaminate the empirical stellar libraries.

We took an iterative approach to precisely determine the redshift of the COS spectrum.
This was necessary because the spectrum is of an off-nuclear position on a steeply rising portion of the rotation curve.
We used the galaxy redshift as a starting point, found the best-fit SB99 continuum model, and then cross-correlated the model with the data (with ISM lines masked out) to find an improved redshift.
The data was then re-fit with the SB99 models and the procedure repeated.

We fit the COS spectrum with a linear combination of SB99 single-age stellar population (SSP) models with ages ranging from 1 -- 20 Myr.
We tested models with different metallicities and ultimately adopted models with a twice-solar abundance because they provide a better fit to the \ion{Si}{iv} stellar wind lines.
Our best-fit model is dominated by a 3~Myr-old stellar population with small contributions from 1, 4, and 15~Myr-old stars.
The mean light-weighed age is 3.4~Myr, slightly younger than the ages measured for heavily embedded star clusters in the nuclear ring \citep[5 -- 6~Myr;][]{Brandl12}.

Our stellar population model fits include the effects of dust using a \cite{Calzetti00} dust attenuation law.
The best-fit stellar continuum reddening is $E(B - V) = 0.64$, approximately 0.1 mag higher than the value found by \citet{Leitherer13} using similar methods.
This small discrepancy in the reddening is indicative of the level of systematic uncertainty.
It largely reflects differences in the assumed star formation history: \citeauthor{Leitherer13} used a SB99 model with a continuous star formation history, whereas we use a sum of SSP models.

Our final best-fit dust-reddened SB99 model is compared to the data in Figure~\ref{fig:absorption}.
The observed spectrum is divided by the model to produce a normalized spectrum which is used for analyzing NGC~7552's ISM lines (see \S\ref{sec:uv_results}).
A final step in our data processing was to fit and remove the ISM absorption arising from the Milky Way.
This was carried out using the procedure described in \citet{Chisholm14}.
Fortunately, the Milky Way ISM lines are blueshifted by approximately 1550~\kms\ relative to the systemic velocity, and thus for the most part do not overlap with absorption associated with NGC~7552's ISM.

\section{Measuring the outflow using optical emission lines}
\label{sec:results}
\begin{figure*}
    \includegraphics[width=0.482\textwidth]{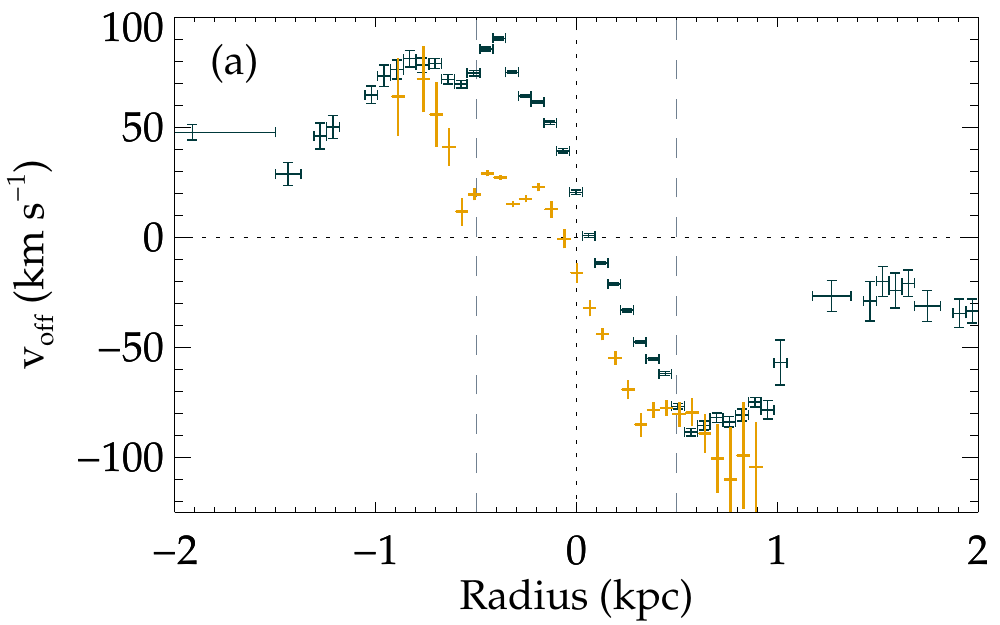}
    ~
    \includegraphics[width=0.482\textwidth]{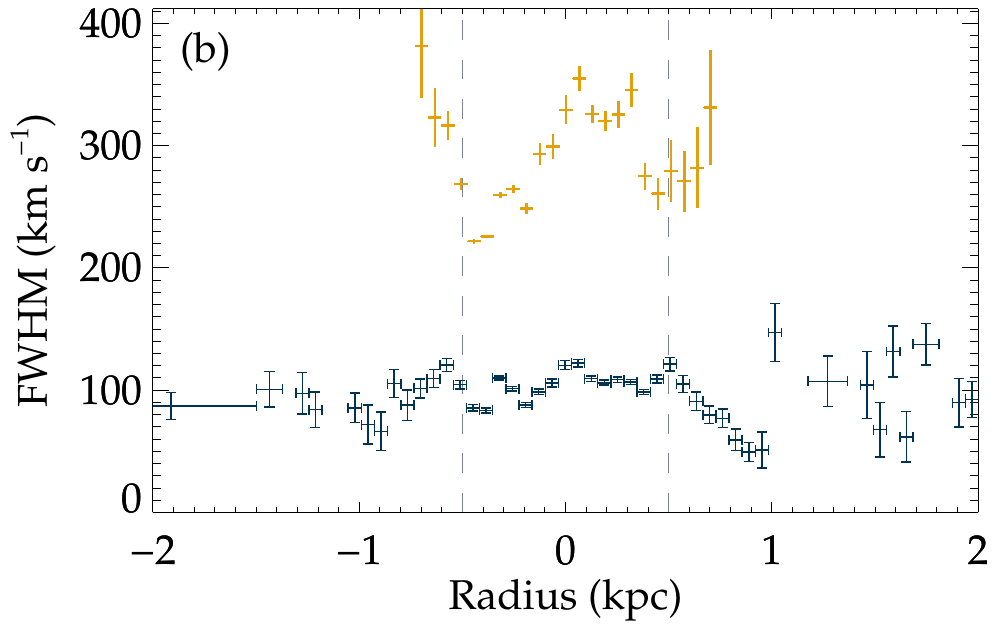} \\
    \includegraphics[width=0.482\textwidth]{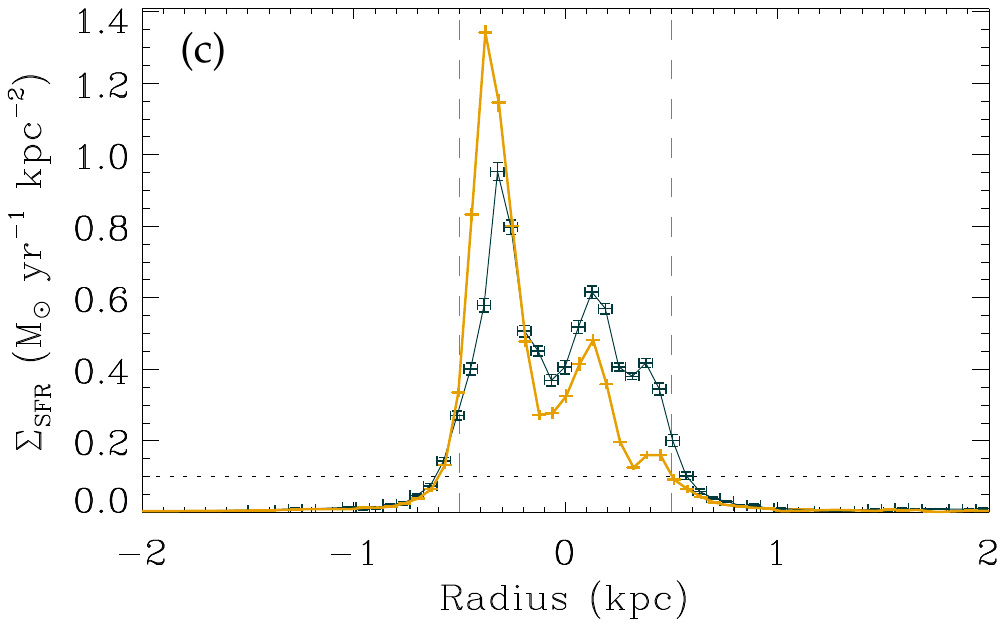}
    ~
    \includegraphics[width=0.482\textwidth]{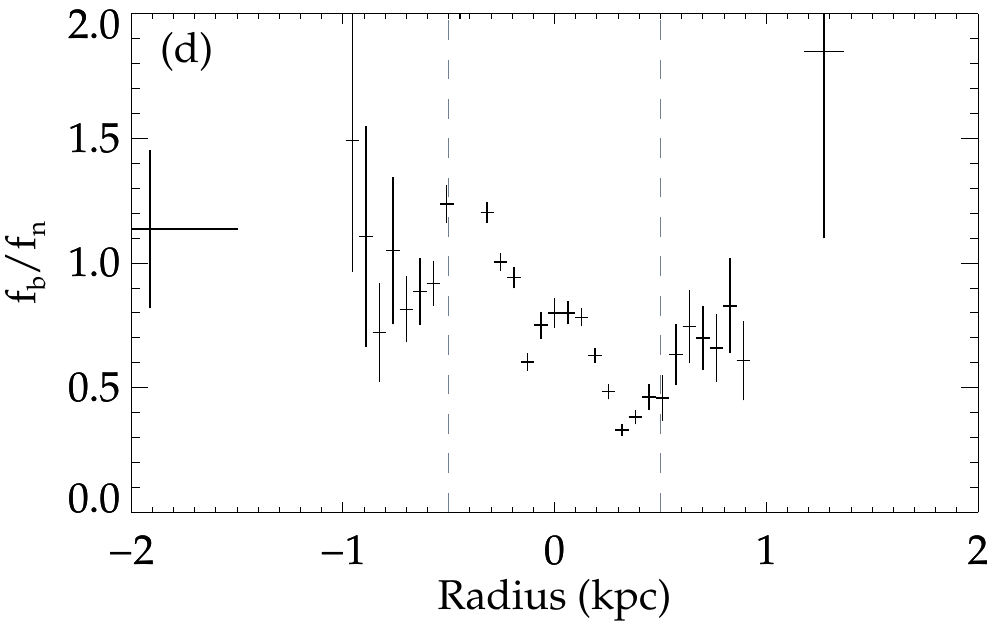} \\
    \includegraphics[width=0.482\textwidth]{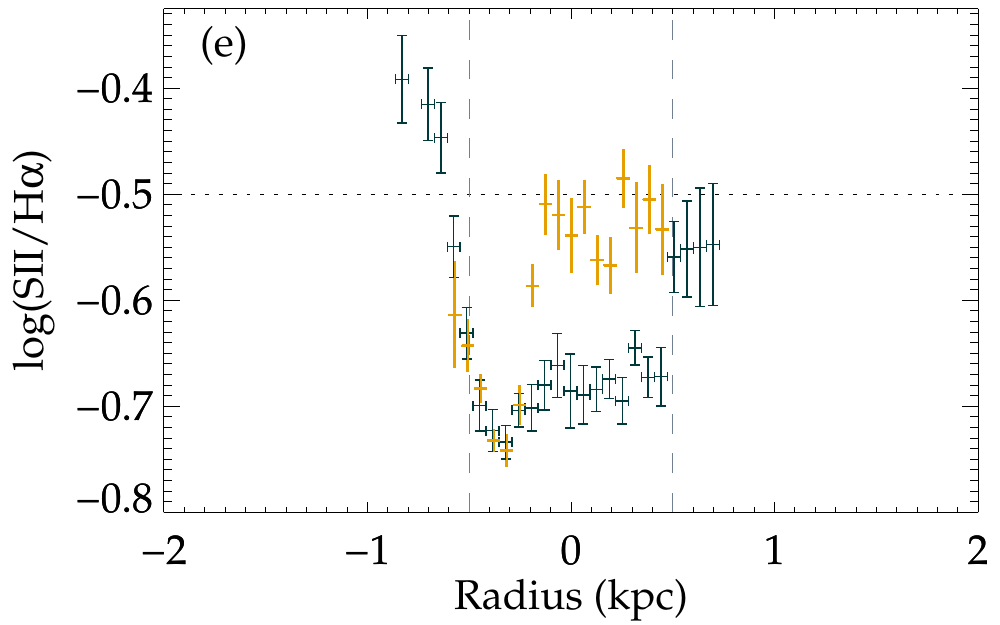}
    ~
    \includegraphics[width=0.482\textwidth]{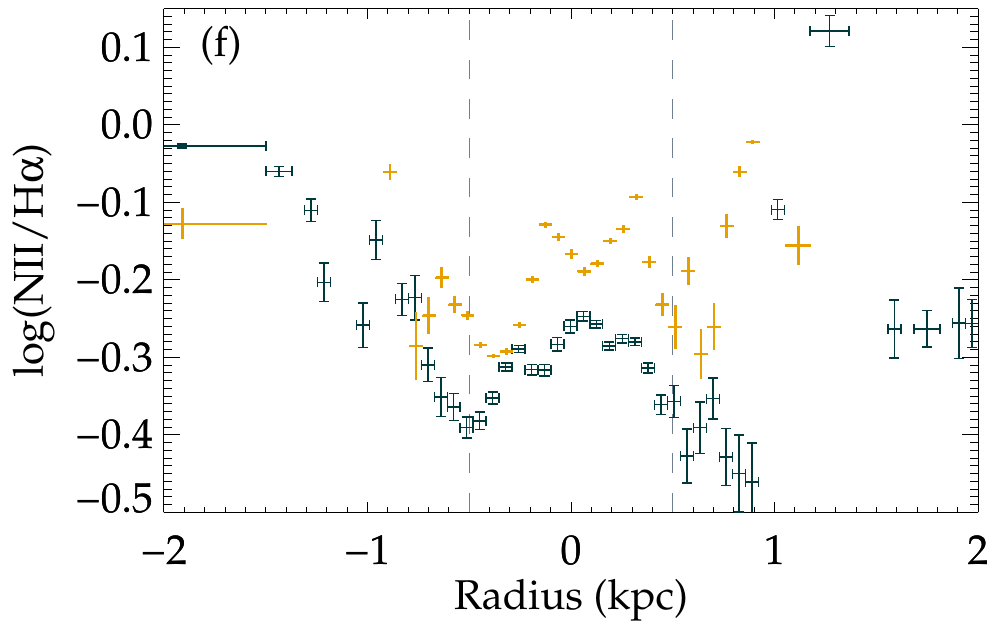}
    \caption{
        Measurements of the kinematics, \Halpha\ line flux, and diagnostic line ratios of NGC~7552.
        Vertical error bars in all figures represent 1$\sigma$ uncertainties, while horizontal error bars represent the radial bin size.
        The vertical dashed lines at $\pm0.5$~kpc in all figures denote the approximate maximum radial extent of the star-forming ring.
        Measurements for all figures are taken at PA $=$ 125$^{\circ}$.
        (a): Line center velocity offsets for the narrow (blue, hats) and broad (orange) velocity components.
        The black dotted lines represent both the velocity and spatial centers of the galaxy.
        (b): Line widths of the narrow (blue, hats) and broad (orange) velocity components.
        (c): Star formation surface density $\Sigma_{SFR}$ measured in the narrow (blue, hats) and broad (orange) \Halpha\ components, using the calibration from \citet{Hao11} without dust correction.
        The horizontal dotted line represents the threshold for driving outflows from \citet{Heckman02}.
        (d): The ratio of broad-line \Halpha\ flux to narrow-line \Halpha\ flux.
        (e): \SII/\Halpha\ line ratios of the narrow (blue, hats) and broad (orange) velocity components.  The horizontal dotted line is the threshold for shocks \citep{Hong13}.
        (f): \NII/\Halpha\ line ratios of the narrow (blue, hats) and broad (orange) velocity components.
        \label{fig:voff}
    }
\end{figure*}

Our fitting routine successfully identifies two emission line components over much of the galaxy, with the most significant detection of broad emission occurring within 1~kpc of the galaxy center.
Representative examples of our extracted emission line data are shown in Figure~\ref{fig:voff}, showing data from PA $= 125\degrees$.
In Figs.~\ref{fig:voff}(a) and (b) we see a clear separation between the narrow and broad components in both the velocity offsets and line widths of the components.
The line centers of the broad components are consistently blueshifted by $\Delta v \approx 30-50$~\kms\ compared to the narrow components, but still largely trace the rotation curve seen in the narrow lines.
The line widths of the broad components also show a distinct separation from the narrow components; we consistently measure line widths of $\sim$300~\kms\ FWHM for the broad components, compared to $\sim$100~\kms\ FWHM for the narrow components.
In Fig.~\ref{fig:voff}(c) we show the star formation surface density $\Sigma_{SFR}$ for both velocity components, calculated from the \Halpha\ luminosity and uncorrected for extinction, using the relation SFR (M$_{\odot}$~yr$^{-1}$) $= 5.53\times10^{-42} L_{\mathrm{H}\alpha}$ (erg~s$^{-1}$) from \citet{Hao11}.
The horizontal dotted line denotes the threshold for driving outflows from \citet{Heckman02}.
Fig.~\ref{fig:voff}(d) shows the ratio of broad \Halpha\ flux to narrow-line \Halpha\ flux, \fbfn.
The broad \Halpha\ flux is consistently at least half that of the narrow flux, and the ratio reaches values above one in regions of highest total \Halpha\ luminosity.
Figs.~\ref{fig:voff}(e) and (f) show the emission line ratio diagnostics $\log$(\SII/\Halpha) and $\log$(\NII/\Halpha).
Overall, we observe larger line ratios in the broad emission line components in both diagnostics with depressed line ratios in regions of highest total \Halpha\ luminosity.

\subsection{Line broadening mechanisms}
\label{subsec:broadening}
As Figure~\ref{fig:voff} shows, the line widths of the narrow components are around 100~\kms\ FWHM ($\sigma \approx 40$~\kms), broad enough to be resolved by our observations.
These line widths are consistent with those seen in \ion{H}{ii} regions of other LIRGs \citep[e.g.,][]{Rich11, Arribas14}, and a factor of $\sim2$ larger than generally seen in local spiral galaxies \citep{Epinat10}.
The spatially-resolved velocity centers of the narrow lines show clear signs of galaxy rotation, further supporting the interpretation that these lines arise from star-forming \ion{H}{ii} regions embedded in the disk of the galaxy.
The broad components have widths of $\sim240-350$~\kms\ ($\sigma \approx$ 100 -- 150~\kms), substantially broader than the narrow components.

Even if NGC~7552 hosts a low-level AGN, the broad line widths are not a result of AGN activity.
Reverberation mapping has shown the diameter of the broad line regions (BLR) of AGN to be much smaller than a parsec \citep{Bentz09, Brewer11}, while we see consistent broad line widths across a region $\sim2$~kpc in diameter.
AGN-driven winds have been invoked to explain spatially-extended broad \Halpha\ emission lines in the central regions of some $z \sim 2$ galaxies \citep{Genzel14}.
However, we consider this explanation unlikely considering the low luminosity of NGC~7552's nuclear source.
The tight correspondence between the broad lines and the inner star-forming ring suggests that the broadening mechanism is related to the star formation process.
It is unlikely that these broad line widths result from typical galactic \ion{H}{ii} regions, and we therefore consider a number of other possible origins.

\subsubsection{A starburst-driven outflow}
\label{subsubsec:outflow}
A number of recent works \citep[e.g.,][]{Genzel11, Soto12, Newman12a, Arribas14} suggest that broad optical emission lines are associated with large-scale galactic winds.
This conclusion is based in part on the finding of predominantly blueshifted line profiles.
And as Figure~\ref{fig:voff}(a) shows, in NGC~7552 the broad line velocity centers are consistently blueshifted by 30 -- 50~\kms\ with respect to the narrow emission across the entire star-forming ring  ($|r| \lesssim 0.5$~kpc).
The ring shows very high levels of extinction with an average $A_V = 5.3$ \citep{Brandl12}, implying that \Halpha\ emission originating on the far side of the disk would be attenuated by a  factor of $\sim 50$.
Thus the blueshifted broad lines in NGC~7552 cannot arise from in-falling material on the far side of the disk; they must be associated with an outflow.

Outflows are frequently modeled as spherical shells of material, moving at a constant velocity $v_{sh}$.
In this simple model, projection effects lead to observed velocities that are smoothly distributed between $-v_{sh}$ and $+v_{sh}$, with a mean velocity of zero and a FHWM of $2v_{sh}$.
This simple model can be adapted to a bi-conical geometry by restricting the angular covering factor of the shell.
When viewed along the axis of the bi-cone, this leads to reduced emission at zero velocity, but the same mean velocity (zero) and FWHM ($2v_{sh}$).
The observed velocities may be affected by extinction associated with the galaxy disk or with the shell itself.
This can lead to a reduction of the flux at positive velocities and a net blueshift of the line profile.
In this physical picture, the observed line width is a consequence of projection effects and extinction, and the most blueshifted gas traces the true deprojected velocity of the shell.  

As we show in Figure~\ref{fig:voff}(a), the broad components of NGC~7552 have FWHM $\sim240-350$~\kms\ and blueshifts of 30 -- 50~\kms\ with respect to the narrow components across the entirety of the star-forming ring.
The simple shell model with extinction has difficulty reproducing these numbers.
If heavy disk extinction removes the redshifted component of the line profile, then the line width is FWHM $\sim v_{sh}$ and the line centroid is $v_{cen} \sim v_{sh}/2$.
Thus, the line width is roughly twice as large as the line centroid velocity: FWHM/$v_{cen}\sim 2$.
In contrast, we measure FWHM$/v_{cen} \sim 7.5$ across most of the star forming ring of NGC~7552.
Thus, we conclude that velocity projection effects associated with a shell-like outflow are not sufficient to explain the broad line profiles;  an additional line broadening mechanism is necessary.

\subsubsection{Turbulent mixing layers}
\label{subsubsec:mixing}
One possible explanation is that the broad lines arise as the result of a turbulent gas mixing layer on the surface of cool gas clumps \citep{Begelman90, Slavin93, Westmoquette07, Tenorio-Tagle10}.
This theory suggests that a hot wind flowing out of the disk and rushing past cool gas clumps induces Kelvin--Helmholtz instabilities on the surface of the clumps; a turbulent mixing layer develops in the boundary region between the hot and cool gas.
The broad lines would in that case be a result of emission from this highly turbulent boundary layer.
This mechanism has been invoked to explain the broad \Halpha\ emission lines (200--300~\kms) seen in the central regions of M82 \citep[e.g.,][]{Westmoquette07}.
One important point is that with the addition of turbulent line broadening, the maximum blueshifted velocity extent of the profile is no longer representative of the bulk flow velocity. 

\subsubsection{Shocks}
\label{subsubsec:shocks}
The large line widths could also be potentially explained as emission from a shocked gas outflow.
In this scenario, bubbles of hot gas, expanding from mechanical energy injected by the starburst, sweep up cool, dense gas in the galaxy, propagating shock fronts into the ISM.
The shocked shells become Rayleigh-Taylor unstable and fragment as they approach the scale height of the disk \citep[][and references therein]{Veilleux05}.
The observed broad line width in this scenario reflects the velocity of the shock \citep{Ho14}. 

Recent studies of U/LIRGs have shown widespread evidence of shocked outflows driven by supernova feedback from star-forming regions.
These works categorize shocked emission lines by comparing nebular emission line ratios of [\ion{O}{iii}]/\Hbeta, [\ion{O}{i}]/\Halpha, \NII/\Halpha, and \SII/\Halpha\ to shock models.
\citet{Rich11} study two local LIRGs and measure broad, shocked emission lines with FWHM $\gtrsim 140$~\kms\ and FWHM $\gtrsim 210$~\kms, respectively, in their two targets.
\citet{Soto12} similarly measure FWHM $> 235$~\kms\ in a sample of 39 ULIRGs.
\citet{Ho14} study an isolated star-forming disk galaxy ($\log M_* = 10.8$~M$_\odot$; SFR $\sim 10$~M$_\odot$~yr$^{-1}$) and find spatially-extended broad velocity components with line ratios that are well-explained by shock models.
They observe a correlation between broad-component line width and the line ratios \NII/\Halpha, \SII/\Halpha, and [\ion{O}{i}]/\Halpha, and find that these correlations are well-explained by shock models where the only free parameter is the shock velocity.

Our observations do not include the [\ion{O}{iii}], [\ion{O}{i}], and \Hbeta\ emission lines, but \citet{Hong13} calibrate shock diagnostics based on single line ratios.
Their results reveal a dividing line between shock ionization and \ion{H}{ii} region photoionization at $\log($\SII/\Halpha$) > -0.5$, with ratios higher than this value being indicative of shock ionization.
Representative examples of the \SII/\Halpha\ and \NII/\Halpha\ line ratios of NGC~7552 are seen in Figures~\ref{fig:voff}(e) and (f), respectively.
In general, the line ratios found in the broad emission lines are $\sim0.15$ dex higher (more shock-like) than those seen in the narrow emission lines, but they fall slightly short of the threshold for shock ionization.
On average, the \NII/H$\alpha$ ratio increases with line with, similar to the trend found by \citet{Ho14}.   
However, NGC~7552 has a much smaller dynamic range in line width ($\sigma=20-150$~\kms) than the galaxy studied by Ho ($\sigma=10-350$~\kms), making detailed comparisons with models more difficult.

A comparison of Figures~\ref{fig:voff}(c), (e), and (f), reveals a strong anti-correlation between \Halpha\ luminosity and the \NII/\Halpha\ and \SII/\Halpha\ nebular line ratios.
As is especially apparent in the \SII/\Halpha\ line diagnostic, the line ratios in the broad and narrow components are approximately equal in regions of very strong \Halpha\ emission (e.g., near $-0.3$~kpc in Fig.~\ref{fig:voff}(e)).
These regions are also where we find the most depressed \SII/\Halpha\ ratios, indicative of photoionization from \ion{H}{ii} regions being the dominant ionization source of the gas.
Curiously, these regions also exhibit the strongest broad-line emission, with the ratio of broad to narrow \Halpha\ emission $f_b/f_n \geq 1$ (Fig.~\ref{fig:voff}(d)).
Outside of these strongest \Halpha\ regions, the narrow-line \SII/\Halpha\ ratio continues to indicate \ion{H}{ii} region photoionization while the broad emission line ratio rises quickly and approaches the shock ionization threshold of $-0.5$, demonstrating the increased contribution of shocks in these regions away from the strongest photoionization.

This anti-correlation between nebular line ratios and \Halpha\ emission strength has been seen observationally in other LIRGs.
\citet{Monreal-Ibero10} find such anti-correlations in integral-field spectra of 32 low-redshift LIRGs.
Their analysis suggests that photoionization from young stars in \ion{H}{ii} regions is sufficient (or nearly-sufficient) in explaining this trend in only isolated, non-interacting systems.
Interacting and merging systems in their study increasingly require the invocation of shocks, however.
These results are relevant to NGC~7552, a LIRG in the interacting Grus Quartet, and serve as further evidence that shocks could be a contributing factor to the nebular line emission we observe.

The anti-correlation between \SII/\Halpha\ and \Halpha\ luminosity in the context of NGC~7552 is also expected since the ages of its dominant nuclear star clusters are 5 -- 6~Myr \citep{Brandl12}.
The ionizing luminosity of an instantaneous starburst dominates the mechanical luminosity until 8~Myr, after which the ionizing luminosity drops sharply while the mechanical luminosity increases until a few tens of Myr \citep[Starburst99,][]{Leitherer99, Sharp10}.
The age of the star clusters in NGC~7552 places them within this region of dominating ionizing luminosity.
Photoionization therefore is likely the primary driver of the nebular line ratios we observe, while shocks in the outflow or turbulent mixing layers could still be responsible for driving the broad line widths.
Higher spatial-resolution spectra could help alleviate this ambiguity.
\citet{Hong13} note that photoionized gas is the dominant gas component in starburst galaxies, and caution that shock-ionized gas is difficult to disentangle from photoionized gas emission at the spatial resolution possible with seeing-limited ground-based observations.

\subsection{Outflow velocities}
\label{subsec:properties}
\begin{figure}
    \includegraphics[width=\columnwidth]{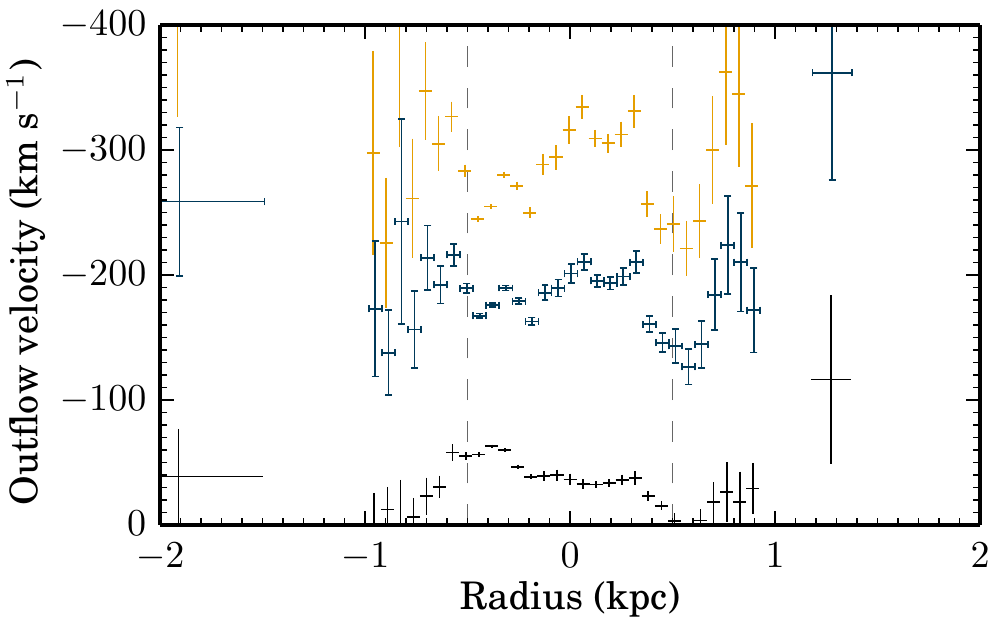}
    \caption{Various measures of outflow velocity at PA $= 125\degrees$.
        Shown are $\Delta v$ (black), the difference between the broad and narrow velocity centers, and the two different measurements of \vmax, \vmaxone\ (blue, hats) and \vmaxtwo\ (orange).
        The vertical dashed lines denote the approximate radial extent of the star-forming ring.
        \label{fig:v_max}
        }
\end{figure}

The net blueshift of the broad \Halpha\ emission lines strongly suggests that the lines arise in an outflow driven by star formation feedback.
However, it is not straightforward to infer the bulk velocity of the outflow from the line profile.
In a simple shell model of the outflow, the maximum blueshifted velocity of the line profile represents the true de-projected velocity of the outflow.
However, we cannot reproduce both the observed line widths and velocity offsets with such a simple model; additional line broadening from shocks or turbulent mixing layers appears to be necessary.
With the addition of these line broadening mechanisms, the maximum blueshifted velocity is no longer indicative of the bulk outflow velocity.
However, for consistency with the literature we proceed with the assumption that $v_{max} = v_{sh}$.
We revisit the issue of the true velocity of the outflow in \S\ref{sec:discussion}.

We can estimate the ``maximum'' outflow velocity of the gas based on the Gaussian line profile of the emission.
Two measures of \vmax\ are summarized by \citet{Westmoquette12}:
$\vmaxone\ \equiv\ v_\mathrm{broad} - \Delta v_\mathrm{broad(FWHM)}/2$ \citep[e.g.,][]{Rupke05, Veilleux05} and  $\vmaxtwo \equiv\ v_\mathrm{broad} - 2\sigma_\mathrm{broad}$ \citep[e.g.,][]{Genzel11}, where $v_\mathrm{broad}$ is the velocity offset of the broad component line center with respect to the narrow component, $\Delta v_\mathrm{broad(FWHM)}$ is the FWHM of the broad component, and $\sigma_\mathrm{broad}$ is the Gaussian sigma value of the broad component, where $\mathrm{FWHM} = 2\sqrt{2\ln2}\,\sigma$.
We refrain from taking the absolute value of \vmaxone\ and \vmaxtwo, as in \citet{Westmoquette12}, to retain clarity about the fact that these velocities are blueshifts.
These measures of \vmax\ are motivated by the fact that the emission is well-characterized by a Gaussian profile.
Eighty-eight percent of the emitting gas is therefore found at velocities red-ward of \vmaxone\ (integrating the profile from \vmaxone\ to $\infty$).
Similarly, 98 percent of the emitting gas is found at velocities red-ward of \vmaxtwo.
In a spherical shell model, \vmax\ represents the deprojected velocity of the outflow.
A comparison of \vmaxone; \vmaxtwo; and $\Delta v$, the velocity difference between the broad line center and the narrow line center; from PA $= 125\degrees$ is shown in Figure~\ref{fig:v_max}.
The velocity offset $\Delta v$ can be assumed to be a lower limit to the bulk outflow velocity of the gas.
We find average values of $\langle\Delta v\rangle = -30 \pm 35$~\kms, $\langle\vmaxone\rangle = -180 \pm 80$~\kms, and $\langle\vmaxtwo\rangle = -290 \pm 100$~\kms, where the stated $1\sigma$ uncertainty is the standard deviation of the distribution of values.

\subsection{Electron density}
\label{subsec:density}
\begin{figure}
    \includegraphics[width=\columnwidth]{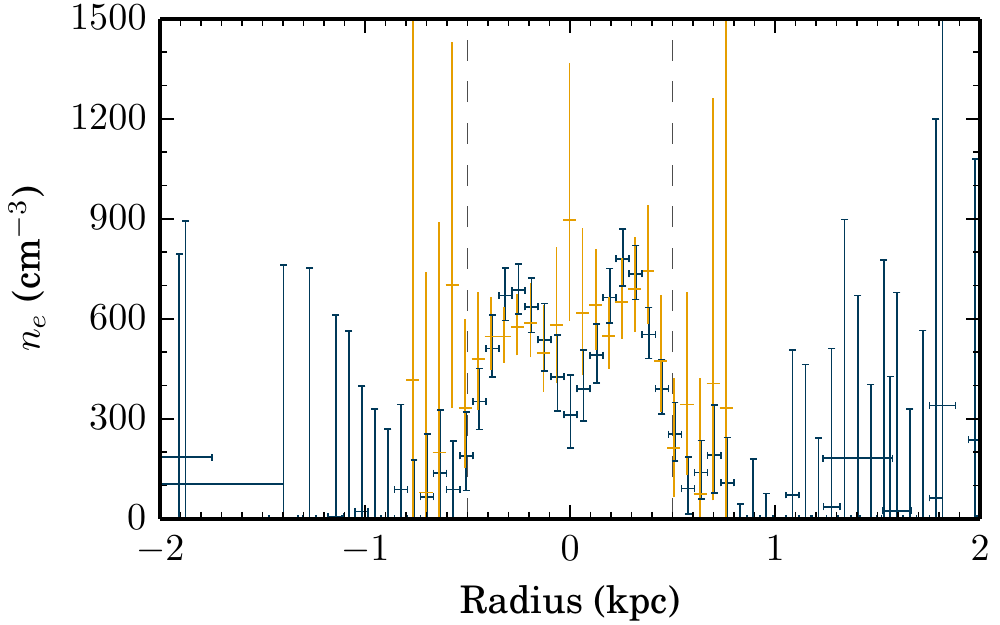}
    \caption{Electron densities measured in the narrow (blue, hats) and broad (orange) components as a function of radius along the slit.  Measurements from all position angles have been averaged together.
        The vertical dashed lines denote the approximate radial extent of the star-forming ring.
        \label{fig:density}}
\end{figure}

Our spectral window includes the \SII\ 6716/6731 line pair, allowing us to calculate the local electron density of the emitting gas.
This quantity is a critical component of the calculation of the mass outflow rate, as will be shown in \S\ref{subsec:mass_outflow}.
We estimate the electron density $n_e$ using the \SII\ 6716/6731 line flux ratio \citep{Osterbrock}, calculated using the \textsc{iraf} \texttt{temden} routine.
The S/N of our \SII\ lines is lower than that of \Halpha\ and \NII, which makes analyzing the \SII\ ratio difficult for individual spectral apertures.
But we can still extract useful information by computing an angular average of the \SII\ measurements from all of our slits, similar to stacking the observations.
By calculating this average we are able to compute separate electron density values for the broad and narrow components, as shown in Figure~\ref{fig:density}.
The error bars are calculated by passing the allowed high and low values of the \SII\ ratio through the \texttt{temden} routine.
The densities measured for the narrow and broad components show very similar values and behavior for $|r| > 0.12$~kpc.
The densities of both components are around $n_e \sim 500-600 \textrm{ cm}^{-3}$ across the star-forming ring, and then drop to $n_e \approx 100 \textrm{ cm}^{-3}$ by $|r| \approx 0.6$~kpc, approximately the low-density limit for this diagnostic.

These densities measured across the star-forming ring are higher than those found in typical Galactic \ion{H}{ii} regions \citep[$\sim 100$~cm$^{-3}$;][]{Osterbrock}.
However, many starburst galaxies show comparable electron densities within their starburst regions.
\citet{Armus89} measure $n_e \approx 350$~cm$^{-3}$ in 33 infrared color-selected LIRGs, and \citet{Arribas14} and \citet{Soto12} measure $n_e \sim 500$~cm$^{-3}$ in two different U/LIRG samples.
In a study of an isolated, local disk galaxy, \citet{Ho14} measure electron densities of a few $\times 100$~cm$^{-3}$ in a broad velocity component (median $n_e = 300$~cm$^{-3}$ in the broadest component), with substantially lower densities in the narrow disk component ($n_e \approx 20$~cm$^{-3}$).
They explain the increased density in the broad lines by invoking compression across shock fronts, lending further credence to the interpretation that broad emission lines originate in shock environments.
We do not see much difference in the densities of our broad and narrow lines.
This may be because the ISM in the high-pressure starburst region of NGC~7552 already has a high gas density and is not as susceptible to further shock compression.
A stack of 14 targets in the \citet{Newman12b} sample have a \SII\ broad-line flux ratio consistent with the low-density limit, and they adopt 50~cm$^{-3}$ for their analysis.
This study is at $z \sim 2$ and therefore may not be representative of the conditions found in starburst/outflow environments in the local universe.

Additionally, we measure somewhat unique behavior in the apertures nearest the nucleus of the galaxy.
Although the narrow-component densities drop steadily toward $R = 0$, which is unsurprising given the ring-like geometry of the dense star-forming region, the densities we measure in the broad component tend to peak toward the center.
The broad-line measurements within $\sim$ 65 -- 125~pc of the center show a break with the narrow-line density profile.
The broad-component density profile shows a strong peak at the nucleus, although the errors on our measurements are also consistent with a flat profile through the nuclear region.

\subsection{Mass outflow rate}
\label{subsec:mass_outflow}
The implications of outflows for galaxy evolution ultimately depend on the mass outflow rate: how much gas is being driven out of the galaxy due to feedback?
To calculate the ionized mass contained within the outflow, we adopt the photo-ionized outflow model from \citet{Genzel11} and \citet{Newman12a}.
The total mass contained within the ionized outflow is given by
\[M_\mathrm{out} = \frac{1.36 m_p}{\gamma_{\mathrm{H}\alpha}n_e} L_{\mathrm{H}\alpha,b}\]
\begin{equation}
    M_\mathrm{out} =
        6.4\times10^6~\textrm{M}_\odot
        \left( \frac{L_{\mathrm{H}\alpha,b}}{10^{42} \textrm{ erg s$^{-1}$}} \right) \!\!
        \left( \frac{500 \textrm{ cm$^{-3}$}}{n_e} \right)
    \label{eqn:mass}
\end{equation}
where $m_p$ is the proton mass, $\gamma_{\mathrm{H}\alpha}$ is the volume emissivity of \Halpha, $n_e$ is the electron density of the outflowing gas, and $L_{\mathrm{H}\alpha,b}$ is the \Halpha\ luminosity in the broad component.
Assuming a gas temperature of $T = 10^4$~K, the volume emissivity of \Halpha\ is $\gamma_{\mathrm{H}\alpha} = 3.56\times10^{-25}$~erg~cm$^{3}$~s$^{-1}$ for Case B recombination \citep{Osterbrock}.
With our eight different slit positions providing good spatial coverage in the inner $\sim 1$~kpc of the galaxy, we are able to calculate the outflow mass in a spatially-resolved fashion.
We calculate the broad-line \Halpha\ luminosity using our adopted distance to NGC~7552 of $D = 22.5 \pm 1.6$~Mpc.
We correct the luminosity for Galactic foreground extinction \citep{Schlafly11}, but \emph{do not} make any corrections for external dust extinction, which we will subsequently address.
Our eight slit positions provide incomplete spatial coverage, so we correct our measured $L_{\mathrm{H}\alpha,b}$ by the fractional spatial coverage of the observations.
We also multiply our measured $L_{\mathrm{H}\alpha,b}$ by 2 to correct for the lack of emission detected from the far side of the galaxy.
For the electron density $n_e$ we adopt the electron density corresponding to the radius of each spectral aperture.
Due to the low S/N of the \SII\ lines, this limits our mass outflow calculation to $r \lesssim 760$~pc, still large enough to encompass the entire projected area of the star-forming ring.
Limiting ourselves to this radius results in 193 independent spectral apertures being used in our mass outflow calculation.
We calculate separate upper and lower uncertainties for our final result by including either the upper or lower uncertainties on $n_e$ in our error propagation.
With these adopted values we calculate a total ionized gas mass of $M_\mathrm{out} = (2.5^{+1.3}_{-0.3}) \times 10^6$~M$_\odot$ in the broad velocity component.

The mass outflow rate is then the total gas mass divided by the outflow timescale, which we approximate as $t_{out} = R_\mathrm{out}/|v_\mathrm{out}|$, the radius of the outflow divided by its velocity.
The ionized mass outflow rate is then
\[\dot{M}_\mathrm{out} = M_{out} / t_{out}\]
\begin{equation}
    \dot{M}_\mathrm{out} = 0.2~\textrm{M}_\odot~\textrm{yr}^{-1}
        \left( \frac{M_\mathrm{out}}{10^6 M_\odot} \right)
        \left( \frac{|v_\mathrm{out}|}{200 \textrm{ km s$^{-1}$}} \right)
        \left( \frac{1 \textrm{ kpc}}{R_\mathrm{out}} \right) .
    \label{eqn:m_dot}
\end{equation}
The radius of the outflow is a fairly uncertain measurement due to the galaxy's face-on orientation.
Other studies typically use the radial extent of broad emission as the value of $R_\mathrm{out}$ \citep[e.g.,][]{Soto12, Rupke13}.
In the same fashion, we detect broad \Halpha\ emission at S/N $\ge 3$ out to approximately 1~kpc in radius, and therefore adopt this value for $R_\mathrm{out}$.

The outflow velocity is also a difficult parameter to estimate accurately.
Numerous studies have used estimates such as \vmaxone\ and \vmaxtwo\ as the outflow velocity, as we have done in order to be consistent with these studies.
These measures depend on the width of the \Halpha\ emission line, which may not be indicative of the wind kinematics as a whole.
If the blue-shifted line centers are more indicative of the bulk gas velocity ($\Delta v \sim 50$~\kms), the calculated mass outflow rate will be smaller by factors of a few.
We investigate the kinematics further in \S\ref{sec:uv_results}.
In order to remain consistent with previous studies of outflows using \Halpha\ emission, we perform two separate calculations of $\dot{M}_\mathrm{out}$ using two different values for $v_\mathrm{out}$: one using $v_\mathrm{out} = \vmaxone$ and one using $v_\mathrm{out} = \vmaxtwo$.
We finally calculate mass outflow rates of $\dot{M}_\mathrm{out} = 0.44^{+0.23}_{-0.04}$~M$_\odot$~yr$^{-1}$ assuming \vmaxone\ and $\dot{M}_\mathrm{out} = 0.70^{+0.37}_{-0.07}$~M$_\odot$~yr$^{-1}$ assuming \vmaxtwo.
For the sake of simplicity, we will adopt only this latter value for our further analysis.

In order to represent this rate by its mass loading factor $\eta$, we assume a star formation rate in the range of $\dot{M}_* =$ 10 -- 15~M$_{\odot}$~yr$^{-1}$, the estimate from \citet{Pan13} of the SFR of the starburst ring.
Assuming the outflow is driven primarily by feedback from the central starburst, it is reasonable to adopt the SFR of that region for our calculations, rather than an estimate of the SFR from the whole disk.
Our estimate of the mass outflow rate corresponds to mass loading factors in the range of $\eta =$ 0.05 -- 0.07 with no extinction correction.

The \citet{Arribas14} study finds average mass loading factors of $\eta = 0.3$ and $\eta = 0.5$, respectively, for the LIRGs and ULIRGs in their sample, with very few systems having $\eta \geq 1$.
These averages are factors of a few higher than our measurements in NGC~7552 but have been corrected for extinction, unlike our data.
Our spectra do not include \Hbeta, so we are unable to do a direct calculation of the broad component extinction.
Without a direct measurement, correcting our data for extinction becomes a much more imprecise task.

Direct measurements of the extinction toward embedded star clusters in the star-forming ring of NGC~7552 are presented by \citet{Brandl12}.
The most embedded clusters have measured extinctions as high as $A_V = 9.0$, with an average value of $A_V = 5.3$ over the star-forming ring.
However, this extinction estimate represents the attenuation by the dense gas within the star-forming ring, along a line-of-sight extending all the way to the deeply-embedded star clusters.
Emission from supernova-driven outflows on the near side of the disk would not suffer from such strong attenuation, and these extinction values are therefore likely to be overestimates.

We also have an estimate of the extinction from our fitting of the UV stellar continuum (\S\ref{subsec:uv}).
Assuming a \citet{Calzetti00} extinction law, this extinction of $E(B - V) = 0.64$ attenuates our \Halpha\ flux by a factor of $7.1$.
That the star cluster along the COS line-of-sight is UV-bright implies that it has broken out of the dense gas in the galaxy's mid-plane and star-forming ring, and the attenuation of the starlight is largely the result of extra-planar gas (e.g., gas entrained in a galactic wind).
Adopting this extinction results in an adjusted mass outflow rate of $\dot{M}_\mathrm{out} = 5.0^{+2.7}_{-0.5}$~M$_\odot$~yr$^{-1}$, or a mass loading factor of $\eta =$ 0.3 -- 0.5, similar to the \citet{Arribas14} study.
We adopt this result as our best estimate of the mass outflow rate.

\section{Measuring the outflow using UV absorption lines}
\label{sec:uv_results}
\begin{figure*}
    \includegraphics[width=\textwidth]{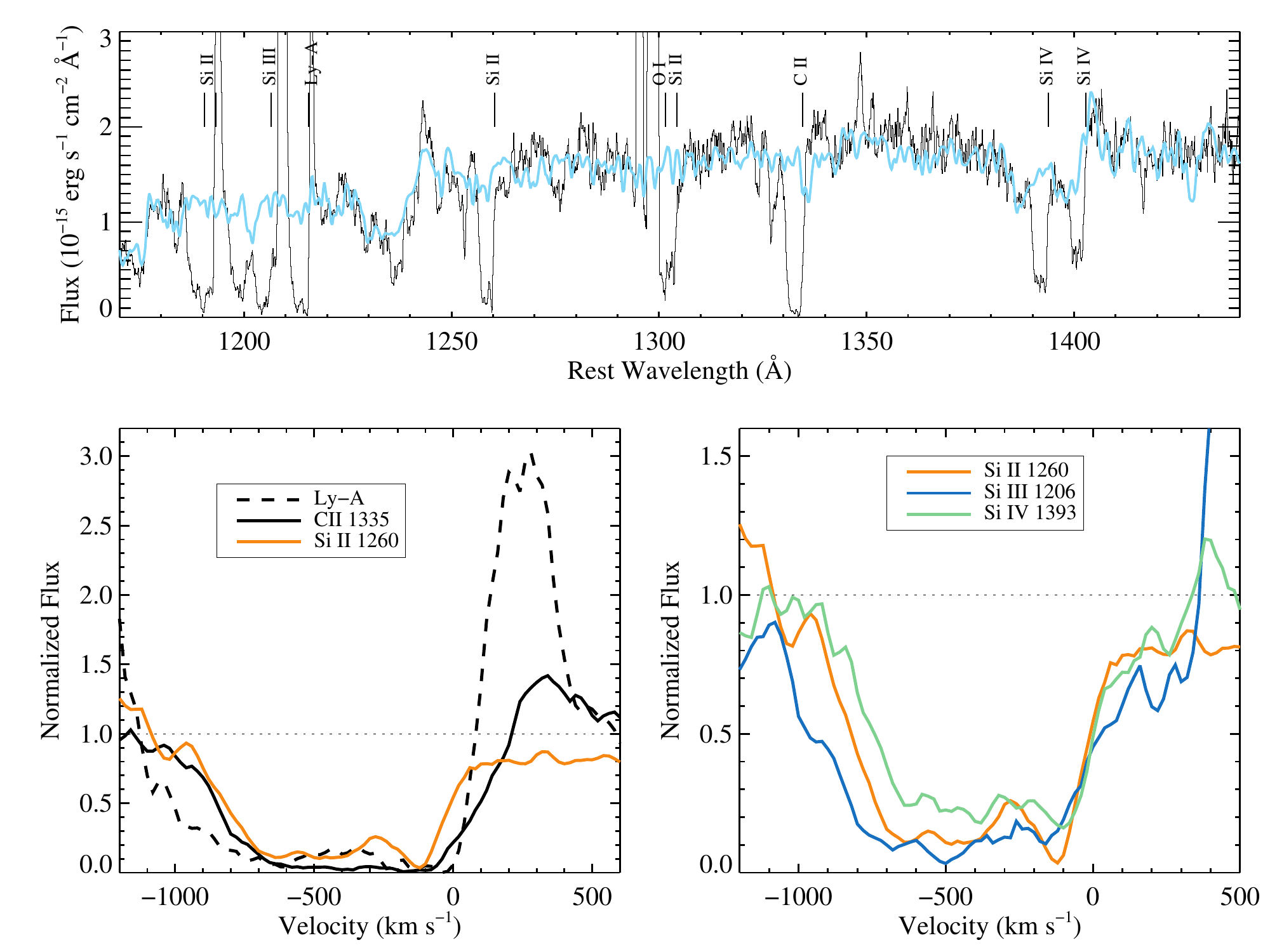}
    \caption{
        \emph{HST}/COS spectrum of NGC~7552.
        The top panel shows the data (black) and a theoretical stellar population model (blue) from STARBURST99.
        Prominent ISM absorption lines are labeled.
        Emission lines other than the marked \Lyalpha\ line are geocoronal lines.
        The lower left panel highlights the continuum-normalized spectra of \Lyalpha, \ion{C}{ii}~1335, and \ion{Si}{ii}~1260.
        These features are some of the strongest and most-saturated transitions in the spectrum, and they show the full extent of the absorption, extending to $>1000$~\kms.
        The absorption profile is very deep (almost reaching zero intensity) from $0$ to $-600$~\kms, suggesting that the gas in the outflow almost completely covers the source.
        In the lower right panel, transitions of \ion{Si}{ii}, \ion{Si}{iii}, and \ion{Si}{iv} are shown.
        The similarity of the line profiles suggests that different ionization phases of the gas have similar spatial distributions and kinematics.
        \label{fig:absorption}}
\end{figure*}
One concern with our analysis is that the gas tracers we have used are only measured in emission, a process that depends on the square of the local gas density.
This has the effect of making our study more sensitive to denser regions of the outflow, and not necessarily representative of the outflow on a large scale.
For this reason it is informative to compare our results to UV absorption-line measurements of NGC~7552.
A study of the mass outflow rate using this data was also presented in \citet{Leitherer13}.
A major difference between that study and our work is that they estimated a total gas column from the measured dust attenuation.
In this study, we pursue an alternative method, using column densities derived from the absorption profiles of Si lines in order to calculate a total mass outflow rate.

\subsection{UV absorption line profiles: gas at 1000~\kms}
In Figure~\ref{fig:absorption} we show the \emph{HST}/COS spectrum of NGC~7552.
The spectrum shows many prominent blueshifted interstellar absorption lines.
To study these lines, we first normalize the spectrum by a theoretical model of the stellar continuum as described in \S\ref{subsec:uv}.
The spectrum and fitted continuum model are shown in the top panel of Figure~\ref{fig:absorption}.
This continuum model accounts for broad absorption from stellar winds that could contaminate our target interstellar lines.
The lower left panel of Fig.~\ref{fig:absorption} shows the line profiles of \Lyalpha, \ion{C}{ii}~1335, and \ion{Si}{ii}~1260.
These are some of the strongest and most saturated transitions in this wavelength range and they are therefore useful for exploring the full velocity extent of the gas associated with the wind.
Absorption is clearly seen in all ions out to 900~\kms\ and in \Lyalpha\ out to 1100~\kms.
These velocities greatly exceed the maximum velocities associated with the \Halpha\ emission ($\Delta v_{max}=160-250$~\kms), a point that we will return to in \S\ref{sec:discussion}.
The absorption is very deep from $v = 0$ to $-600$~\kms, almost reaching zero intensity.
This implies that the gas in the wind at these velocities almost completely covers the UV-luminous stars.

The \Lyalpha\ line shows a relatively typical profile for a star forming galaxy: redshifted emission and blueshifted absorption.
Such a profile can arise when \Lyalpha\ photons resonantly scatter in the wind.
The redshifted emission is thought to be produced by photons scattering off the opposite side of the wind (e.g., the portion of the outflow moving away from the observer).
It is somewhat surprising that we detect \Lyalpha\ emission in our COS observations of NGC~7552 since we are observing a UV-bright cluster complex against a backdrop of very dusty, heavily-embedded clusters.
\citet{Brandl12} compute an average dust attenuation for NGC~7552's star forming ring of $A_V = 5.3$, which corresponds to attenuation by more than a factor of a million at 1216~\AA.
Since the ring is effectively opaque in the UV, it is unlikely that the \Lyalpha\ emission comes from the far side of a galaxy-scale outflow.
It is more plausible that the emission arises from the back side of a cluster-scale outflow (e.g., from gas on the near side of the disk that is being driven away from the observer and towards the disk mid-plane.)
\citet{Leitherer13} model the \Lyalpha\ profile of NGC~7552 using the radiative transfer models of \citet{Schaerer11}.
They find a good fit to the emission profile using a spherical shell model with a constant expansion velocity of 50~\kms, which is comparable to the velocity of the broad \Halpha\ component.
However, the model substantially under-predicts the high velocity \Lyalpha\ absorption.
One plausible explanation is that the absorption traces both a low velocity cluster-scale outflow and a higher velocity galactic-scale outflow.
We consider this idea further in section \S{\ref{sec:discussion}.
 
\subsection{Measuring column densities}
 \begin{figure*}
    \includegraphics[width=\textwidth]{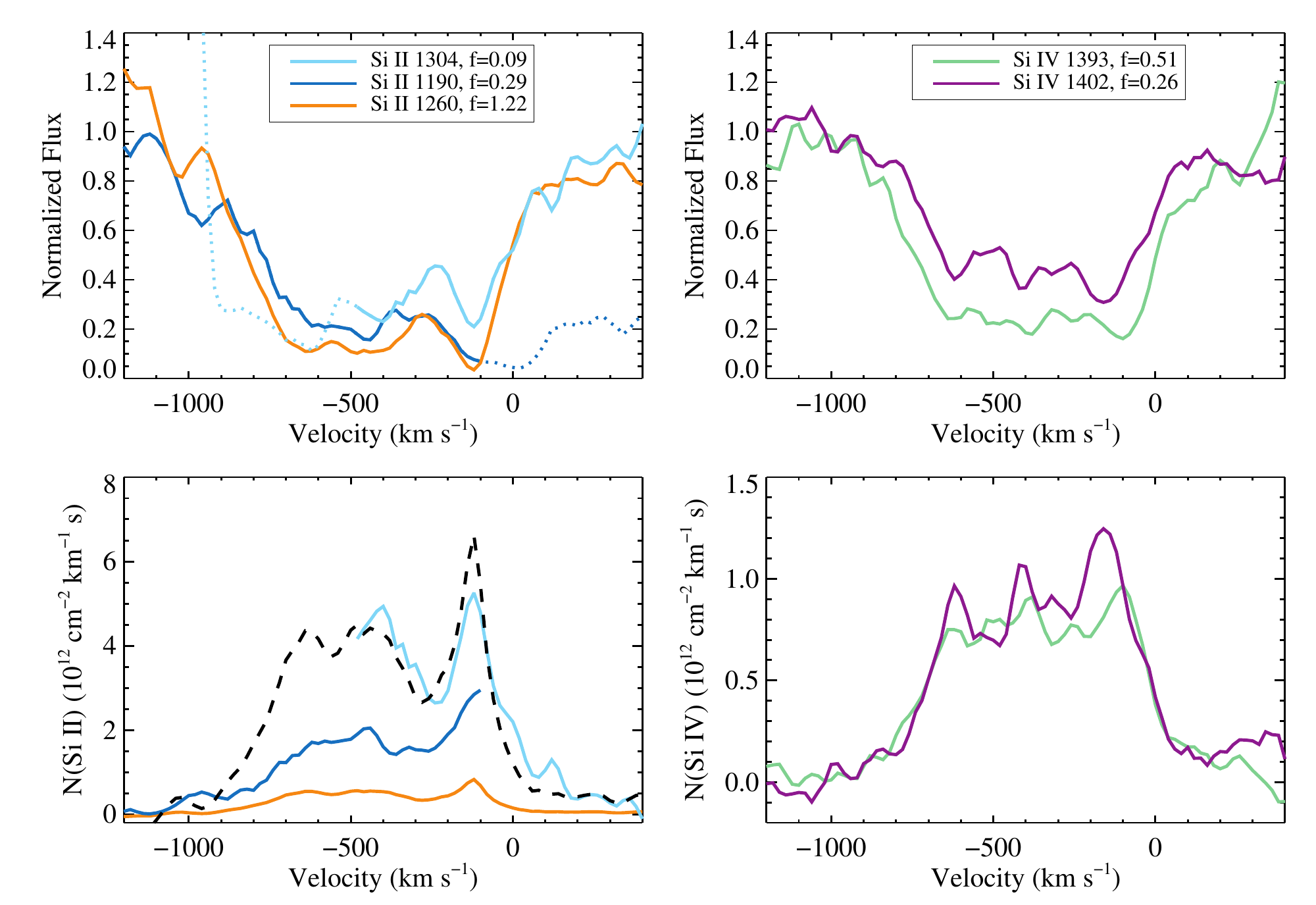}
    \caption{
        Line and column density profiles derived from various \ion{Si}{ii} and \ion{S}{iv} transitions.
        The transitions and their oscillator strengths are listed in the legend.
        Portions of the line profile that are contaminated by other lines are shown as dotted lines.
        The disagreement between the derived column densities of \ion{Si}{ii} may be due to a non-unity covering factor or the presence of unresolved saturated components.
        The low $f$-value transition of \ion{Si}{ii}~1304 is likely to yield the most reliable column density, however the blue portion of the profile is contaminated by \ion{O}{i}~1302.
        The dashed black line shows the column density derived from \ion{Si}{ii}~1260 arbitrarily scaled up by a factor of 8.
        Since this provides a good match to the column derived from \ion{Si}{ii}~1304, we use this to estimate the total \ion{Si}{ii} column.
        The \ion{Si}{iv} lines are not strongly saturated and yield column densities that are in reasonable agreement.
        \label{fig:Si_column}}
\end{figure*}

Our goal is to use the UV lines to estimate the column density of the outflowing gas and, ultimately, to compute the mass outflow rate.
The first step is to derive accurate ionic column densities for all of the important \ion{Si}{ii} phases and combine them to arrive at a total Si column.
By applying a correction for depletion onto dust and assuming a solar Si/H ratio, we can estimate the total hydrogen column.
The right-hand panel of Fig.~\ref{fig:absorption} shows transitions of \ion{Si}{ii}, \ion{Si}{iii}, and \ion{Si}{iv}, which have ionization potentials of 8.15, 16.35, and 33.49~eV, respectively.
The overall shapes of the line profiles are very similar, suggesting that all of the lines are associated with the warm neutral phase of the ISM.
Unfortunately, the strongly saturated nature of the \ion{Si}{ii}~1260 and \ion{Si}{iii}~1206 lines makes it difficult to derive reliable ionic column densities.

In Figure~\ref{fig:Si_column} we show column densities derived using the apparent optical depth method \citep{SavageSembach91} for the \ion{Si}{ii}~1190, \ion{Si}{ii}~1260, and \ion{Si}{ii}~1304 lines, which have $f$-values that differ by as much as a factor of 14.}
In the ideal case, all three transitions would yield the same column density.
The fact that the lines with lower $f$-values have higher derived column densities may indicate that unresolved saturated components are present \citep{SavageSembach91} or that the covering factor of the gas is less than unity.
The low $f$-value transition of \ion{Si}{ii}~1304 is likely to yield the most reliable column density, however the blue portion of the profile is contaminated by \ion{O}{i}~1302 geocoronal emission.
We find that scaling up the column density derived from the strong unblended \ion{Si}{ii}~1260 line by a factor of 8 provides a reasonably good match to the column derived from \ion{Si}{ii}~1304, and we use this rescaled line profile to estimate the total \ion{Si}{ii} column.
The sole \ion{Si}{iii} transition $\lambda1206$ has a high oscillator strength and is likely strongly saturated.
The \ion{Si}{iv}~1393, 1402 lines do not appear to be strongly saturated and yield column densities that are in reasonable agreement with one another.

The ratio of \ion{Si}{ii} to \ion{Si}{iv} is approximately 5:1.
Assuming that the gas arises from a single phase, we can use the \ion{Si}{ii}/\ion{Si}{iv} ratio to estimate the contribution of \ion{Si}{iii} to the total Si column.
We compare our measured ratio to the photoionization model predictions for circum-galactic gas presented in \citet{Werk14}.
We use the model shown in their Fig.~41 which is for log~N(\ion{H}{i}) = 17.
In the model, the relative ratios of different ions depend upon the ionization parameter.
We constrain the ionization parameter using our measured ratio of \ion{Si}{ii}/\ion{Si}{iv} and infer N(\ion{Si}{iii}) $=  2 \times$ N(\ion{Si}{ii}) from this model.
This photoionization model was designed for circum-galactic gas illuminated by the diffuse UV background, and therefore may not be an accurate representation of the gas in the inner parts of the galactic wind, which is likely a complex, multi-phase structure.
Additionally, this model assumes a metallicity of $\log(Z/Z_\odot) = -0.4$, likely lower than the metallicity of NGC~7552.

We compute the total \ion{Si}{ii} column as a function of velocity by summing the \ion{Si}{ii}, \ion{Si}{iii}, and \ion{Si}{iv} profiles and find N(Si) $= 9.86 \times 10^{15}$~cm$^{-2}$.
We turn the Si column into an \ion{H}{i} column by applying a correction for dust depletion and assuming a solar Si/H ratio, using values for these parameters of D(Si) $=-0.29$ and log(Si/H)$_{\sun}=-4.45$ from \citet{Draine11}.  The total hydrogen column density in the outflow, estimated from the \ion{Si}{ii}, \ion{Si}{iii}, and \ion{Si}{iv} absorption lines between velocities of 0 and $-1000$~\kms, is $N(H) = 5.28 \times 10^{20}$~cm$^{-2}$.

One disadvantage of the apparent optical depth method is the implicit assumption that the gas completely covers the source.
In principle, one can obtain information about \emph{both} the covering factor and the column density by simultaneously analyzing multiple transitions of a given ion that span a range of $f$-values.
We explore this approach by simultaneously fitting the \ion{Si}{ii}~1190,1193; \ion{Si}{ii}~1260; and \ion{Si}{ii}~1304 lines with Voigt profiles.
The observed line profiles are complex and clearly contain multiple blended components.
Given our modest spectral resolution ($\sim100$~\kms), we model each absorption feature with a set of seven individual velocity components.
Each velocity component is fit with a Voigt profile with the $b$-parameter set to half the instrumental resolution.
The covering factor and \ion{Si}{ii} column density are free parameters for each component.

The similarity of the \ion{Si}{ii}, \ion{Si}{iii}, and \ion{Si}{iv} line profiles (see Fig.~\ref{fig:absorption}) motivated us to include the \ion{Si}{iii} 1206, and \ion{Si}{iv} 1393, 1402 lines in our fits.
We allow the ionic ratios (e.g., \ion{Si}{iv}/\ion{Si}{ii}) to vary, but we assume that this ratio is the same for all velocity components.
With these assumptions, we achieve excellent fits to all of the individual absorption line profiles.
Once again, the \ion{Si}{iii} column is poorly constrained; we therefore assume N(\ion{Si}{iii}) $ = 2\times$ N(\ion{Si}{ii}) as we did previously.
Summing over all blueshifted components we find $N(\mathrm{Si}) = (1.71 \pm 0.45) \times 10^{16}$~cm$^{-2}$ which yields $N(H) = (9.17 \pm 2.41) \times 10^{20}$~cm$^{-2}$ using the same abundance and dust depletion corrections as above.
This column is roughly a factor of two larger than we derived using the apparent optical depth method, but is consistent with that measurement within the errors.
Most of this increase comes from a much larger derived column at low velocities ($< 200$~\kms).

\citet{Leitherer13} estimate the hydrogen column density in a different manner, using the galactic relationship between N(H) and reddening from \citet{DiplasSavage94}
\begin{equation}
    \textrm{N(H)} = 4.9 \times 10^{21}~E(B - V)~\textrm{cm}^{-2} .
\end{equation}
Using the $E(B - V) = 0.64$ derived from fitting the stellar continuum (\S\ref{subsec:uv}), we calculate $\textrm{N(H)} = 3.1 \times 10^{21}~\textrm{cm}^{-2}$, about a factor of 5.9 and 3.4 higher than the value derived from the Si absorption lines using the two methods described above.
It is reasonable to expect that this value represents an upper limit on the column density, as the extinction traces the column to the star cluster and thus likely includes a contribution from both gas in the disk and the wind.
To be maximally conservative, we adopt the lowest of our three \ion{H}{i} column density estimates, $N(H) = 5.28 \times 10^{20}$~cm$^{-2}$, derived using the apparent optical depth method.

\subsection{The wind mass and mass outflow rate}
\label{subsec:uv_mout}
\begin{figure}
    \includegraphics[width=\columnwidth]{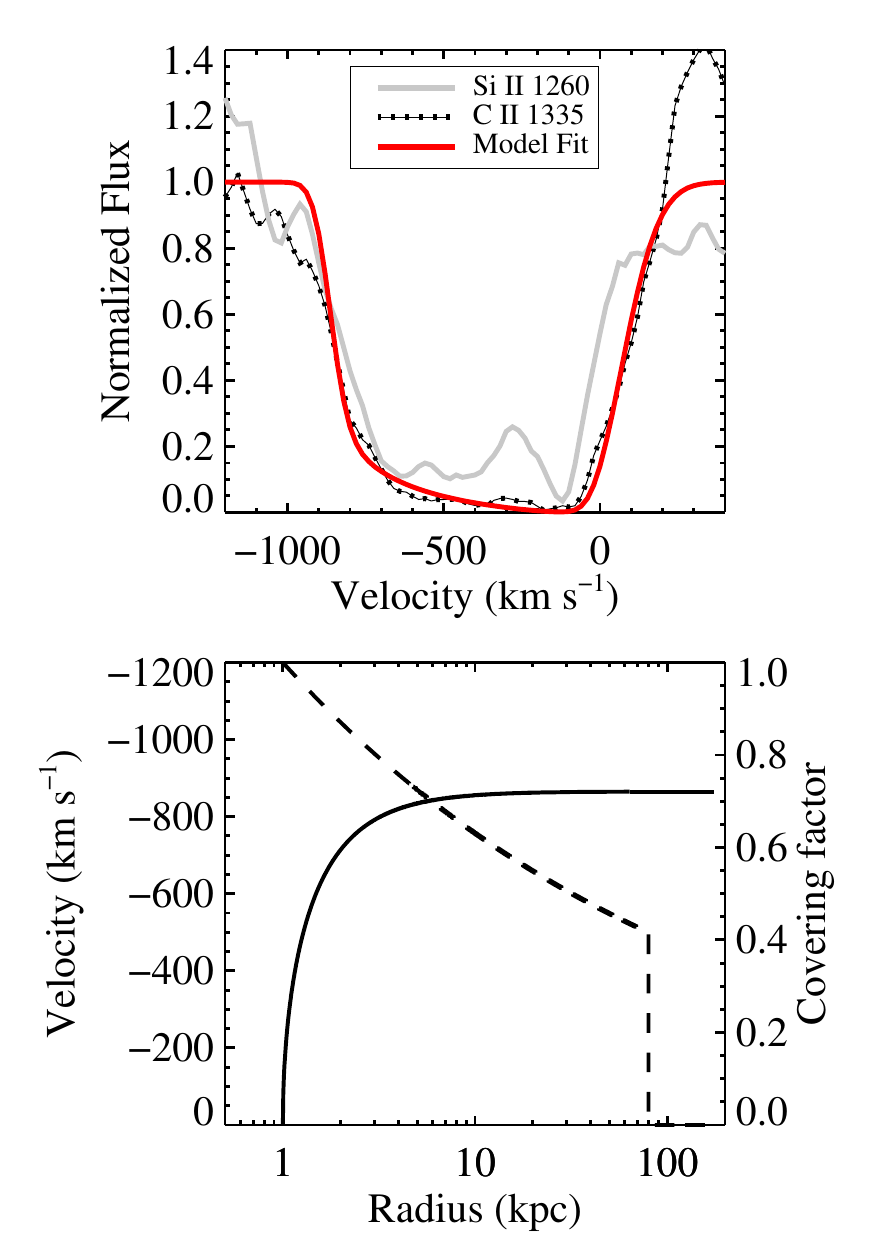}
    \caption{Outflow model fit to NGC~7552's UV absorption lines.
        The model, from \citet{Steidel10}, requires both the cloud covering factor and acceleration to have a power law dependence on the distance from the starburst.
        In the top panel, the gray and black lines show the \ion{Si}{ii}~1260 and \ion{C}{ii}~1335 absorption lines and the red line is the best fit model.
        The bottom panel shows the resulting dependence of the velocity (solid) and cloud covering factor (dashed) on radius.
        Note that most of the gas acceleration happens over a very small range of radii (1 -- 3~kpc).
        The minimum and maximum radius (1 and 100~kpc) are not free parameters of the fit.
        \label{fig:steidel}}
\end{figure}

Following \citet{Rupke05}, we model the outflow as a thick shell with inner and outer radii $r_1$ and $r_2$.
The total mass of the outflow is computed as
\begin{equation}
M_{wind} = 4\pi C_{\Omega} C_f \mu m_p N(H) r_1 r_2 .
\end{equation}
Here $C_{\Omega}$ represents the angular covering fraction of the wind, $C_f$ the clump covering factor, $\mu$ the mean molecular weight, and $m_p$ the proton mass.
We adopt $C_{\Omega} = 0.67$, assuming that the wind has a bi-conical geometry with a half-opening angle of $60^{\circ}$ \citep{Chen10}.
(Note that \citet{Rupke05} assume a slightly smaller value of $C_{\Omega} = 0.4$ for LIRGs).
The clump covering factor $C_f$ accounts for the fact that the wind contains discreet clouds that do not completely cover the UV emitting stars.
Based on our multi-component fits to the line profiles we find that the covering factor ranges from 0.6 -- 1.
We adopt the average value, $C_f = 0.9$, in our calculation.
We assume $\mu = 1.4$ to account for both hydrogen and helium.
The outflow radii are the least well-constrained parameters of the equation.
For the inner radius we adopt $r_1 = 1$~kpc, the radius we assumed for our \Halpha\ emission in \S\ref{subsec:mass_outflow}.
Given that the velocity extent of the UV absorption is much larger than that of the \Halpha\ emission, it is reasonable to assume that the outer radius of the shell is substantially larger than that traced by \Halpha.
\citet{Chisholm14} adopt a wind radius of 4.4~kpc based on an observed scaling between star formation and outflow extent \citep{Grimes05}.
We adopt this value as our outer radius $r_2$.
\citet{Rupke05} and other authors, including \citet{Leitherer13}, frequently assume $r= 5$~kpc, although these particular studies use thin shell models ($r_1 \approx r_2$).
Rewriting the above formula in more convenient units and plugging in values we find
\begin{eqnarray}
    M_{wind} = 2.0 \times 10^8~\textrm{M}_{\sun}
    \left(\frac{C_{\Omega}}{0.67} \right)
    \left(\frac{C_f}{0.90}\right) \times \nonumber\\
    \left(\frac{N(H)}{5.28 \times 10^{20}~\textrm{cm}^{-2}}\right)
    \left(\frac{r_1}{1~\textrm{kpc}} \right)
    \left(\frac{r_2}{4.4~\textrm{kpc}} \right) .
\end{eqnarray}
To compute the average mass outflow rate, we divide the wind mass by the average timescale of the wind, $t=r_2/v$, where $v$ is the average outflow velocity and $r_2$ is the outer shell radius.
Here we have adopted the column-density-weighted average of $v = 410$~\kms, which yields a timescale of 10~Myr.
Plugging in we find
\begin{eqnarray}
    \dot{M}_{wind} = 19~\textrm{M}_{\sun}~\textrm{yr}^{-1}~
    \left(\frac{C_{\Omega}}{0.67} \right)
    \left(\frac{C_f}{0.90}\right) \times \nonumber\\
    \left(\frac{N(H)}{5.28 \times 10^{20}~\textrm{cm}^{-2}}\right)
    \left(\frac{r_1}{1~\textrm{kpc}} \right)
    \left(\frac{v}{410~\textrm{\kms}} \right) ,
\end{eqnarray}
implying a mass loading factor in the range $\eta = 1.3$ -- 1.9. 
This contrasts with the \citet{Leitherer13} study, which estimates mass outflow rates on the order of $\sim 100$~M$_\odot$~yr$^{-1}$ for their sample.
Their result could be an over-estimate, given that the dust attenuation is measured in the starlight, while the high-velocity gas likely exists at larger disk heights.
There is also significant uncertainty in our measurements given the large ionization corrections for the various Si species.

In practice, it is likely that the outflow spans a range of radii.
\citet{Steidel10} develop a model of the kinematics and geometry of circumgalactic gas based on extensive observations of UV absorption lines in projected galaxy pairs at $z = 2 - 3$.
In their model, cool dense clouds are embedded in a hot, diffuse, volume-filling medium.
These clouds are accelerated over time by radiation pressure and ram pressure from the hot wind.
As they move outward, the external pressure supplied by the hot wind declines and the clouds expand.
But they do not expand by enough to counteract their geometrical dilution, and thus the cloud covering factor drops with radius as $C_f \propto r^{- \gamma}$, with $\gamma = 0.2-0.6$.
For the Lyman Break galaxies modeled by Steidel, typical maximum values of the velocity and radius are $v_{max} \approx 750$~\kms\ and $r_{max} \approx 80$~kpc.
This outer wind radius is significantly larger than the radius we have assumed ($r = 4.4$~kpc).
We have experimented with fitting Steidel's analytic model to the absorption profile of NGC~7552\ in order to determine the relationship between velocity and radius.
However, we were unable to produce a satisfactory fit due to the fact that the cloud covering factor in NGC~7552\ is roughly constant over a huge range of velocity ($v=0$ to $-600$~\kms).
As shown in Figure~\ref{fig:steidel}, the best fitting model requires a very rapid acceleration of the gas over a very small range of radius (1--3~kpc), which may be difficult to achieve.
Unfortunately, the model has many free parameters and it is not possible to uniquely constrain the inner and outer radius of the outflow (e.g., equally good fits can be obtained by setting the starting radius to 0.1~kpc or the outer radius to 10~kpc).
What is clear is that it is a challenge to devise a physical model to explain how the cloud covering factor remains roughly constant over a very large range of velocity.

\section{Discussion}
\label{sec:discussion}
\begin{figure}
    \includegraphics[width=\columnwidth]{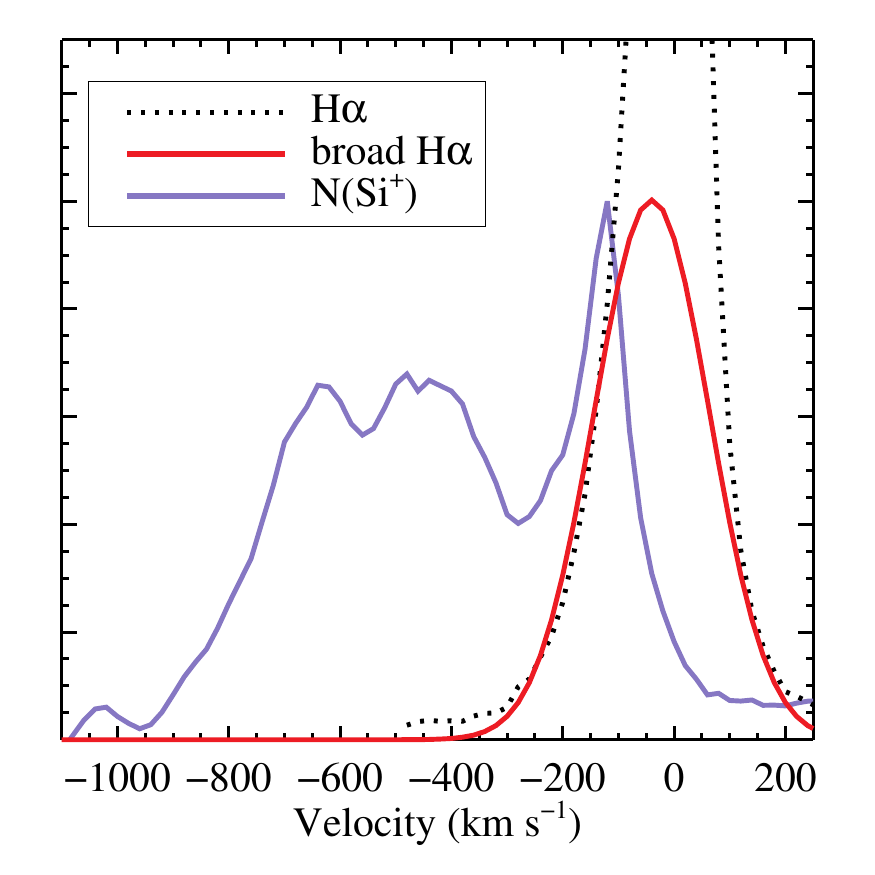}
    \caption{
        Comparison of the UV absorption and \Halpha\ emission showing the ionized silicon column density inferred from \ion{Si}{ii}, \ion{Si}{iii}, and \ion{Si}{iv} UV absorption lines (periwinkle), the coincident \Halpha\ profile (black dotted), and the fit to the broad \Halpha\ velocity component (red).
        The \Halpha\ data has been truncated at 500~\kms\ due to contamination from the \NII~6548 line.
        It is important to note that the Si absorption traces only gas in front of the star cluster along our line-of-sight, while \Halpha\ is sensitive to gas both in front and behind the cluster, largely dependent on only extinction effects.
        \label{fig:uv_vs_ha}}
\end{figure}

We have measured the outflow mass and mass outflow rate of the nuclear starburst region of NGC~7552 using both \Halpha\ emission and UV absorption lines.
For the \Halpha\ line, we measure the mass outflow by decomposing the visible spectrum into broad and narrow velocity components.
We detect broad-line emission along all sightlines within $\sim 1$~kpc of the galactic nucleus.
The narrow-line emission shows a reasonably well-behaved rotation profile.
The broad emission also largely traces this rotation pattern, but is blueshifted by $\Delta v \approx$ 30 -- 50~\kms.
We find that the broad component has line widths of 240 -- 350~\kms\ FWHM, while the line widths in the narrow component are $\sim$100~\kms.
Our measurements of the narrow component are consistent with it arising from star-forming \ion{H}{ii} regions within the disk.
We attribute the broad component to a feedback-driven galactic wind, similar to recent studies \citep[e.g.,][]{Newman12b}, and calculate an ionized mass outflow rate from the broad \Halpha\ flux.
Combined with a measurement of the local electron density $n_e$ via the \SII\ line ratio diagnostic, this measurement yields a mass outflow rate of $\dot{M}_{out} = 5.0^{+2.7}_{-0.5}$~M$_\odot$~yr$^{-1}$, with a mass loading factor of $\eta =$ 0.3 -- 0.5.

Our measurements show electron densities of $n_e \gtrsim 500$~cm$^{-3}$ in the broad lines, substantially higher than what has been measured or assumed in some recent studies (typically $n_e \approx$ 10 -- 50~cm$^{-3}$) \citep{Newman12b, Rupke13}.
We note, however, that the measurement of $n_e$ by \citet{Newman12b} ($n_e = 10^{+590}_{-10}$~cm$^{-3}$) is still statistically consistent with our measurements, considering the large uncertainty in their value.
That study also measures $n_e$ using a spatially-integrated stack of 14 galaxies, and the resulting \SII\ ratios may be sensitive to regions outside the main starburst of these systems.
Many local studies of (U)LIRG/starburst systems find elevated electron densities ($n_e \gtrsim 300$~cm$^{-3}$) in the central star-forming regions \citep[e.g.,][]{Armus89, Lehnert96, Weistrop12, Arribas14}, as well as large declines in $n_e$ with radius.
These high central densities and radial dependences suggest measuring $n_e$ in a spatially-resolved fashion may be important for constraining the densities of the starburst region.
Given that Eqn.~\ref{eqn:mass} depends inversely on $n_e$, increasing the precision in measuring $n_e$ could have order-of-magnitude effects on the measurement of $\dot{M}_\mathrm{out}$.
This may explain the largest source of the discrepancy between our measurements and those at high-redshift, although it is not clear that winds at high redshift are analogous to local starburst-driven winds.

\begin{figure*}
    \centering
    \includegraphics{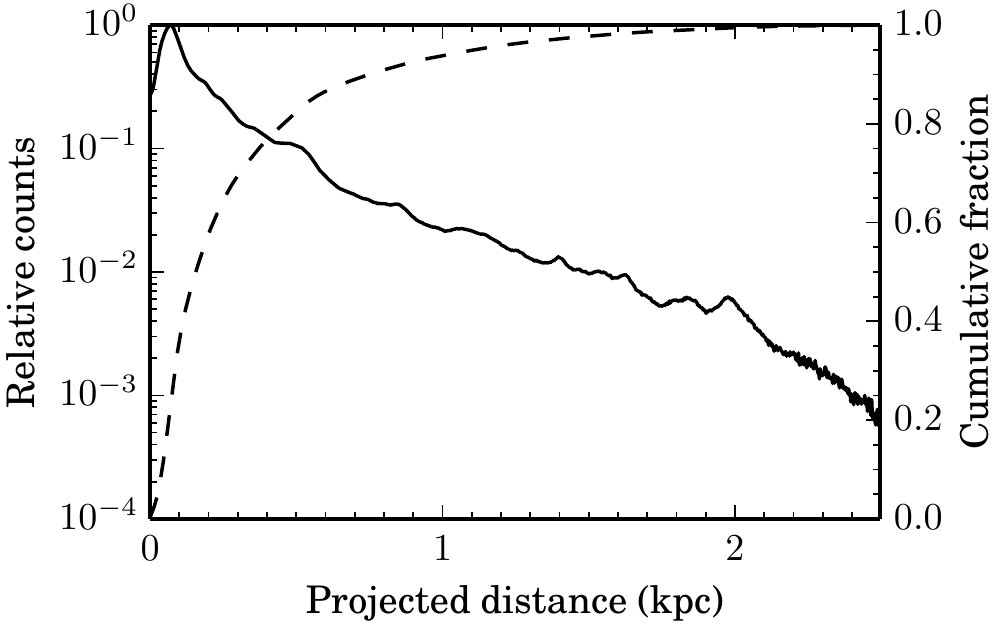}
    \includegraphics{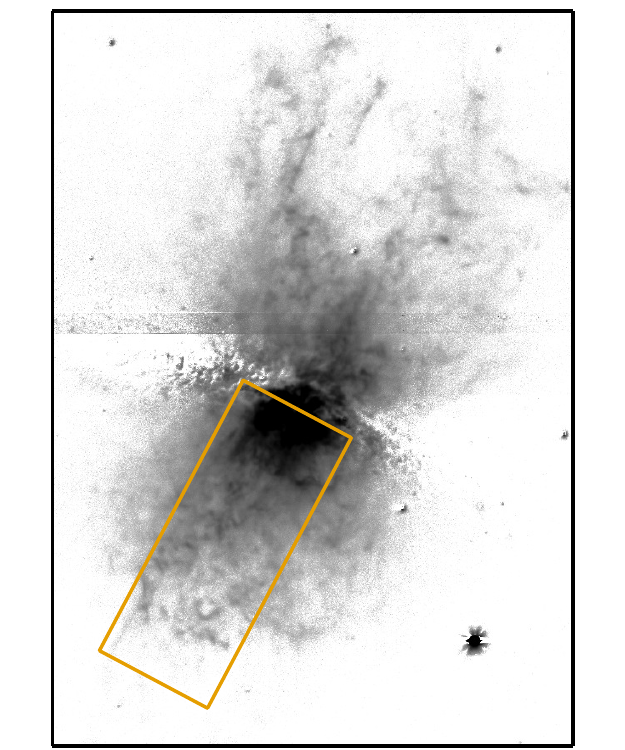}
    \caption{
        \emph{Left:}
        Log-scaled (solid) and cumulative fraction (dashed) of \Halpha\ counts in M82, as a function of projected height above the disk.
        Counts are summed within a 1~kpc-wide box centered about the nucleus.
        \emph{Right:}
        Continuum-subtracted \Halpha\ grayscale image of M82 \citep{WestmoquetteThesis}.  Image is oriented with north up and east to the left.
        The orange box represents the $1 \times 2.5$~kpc extraction region.
        \label{fig:m82}}
\end{figure*}

A main source of uncertainty in our calculation of $\dot{M}_\mathrm{out}$ is extinction in our observations because our spectra contain no reliable method for estimating the extinction.
Our spectral window does not include \Hbeta\ and we are therefore unable to directly measure the extinction via the Balmer decrement \Halpha/\Hbeta.
Our best estimate of the extinction in the broad \Halpha\ component is made by using the reddening of starlight measured via our UV stellar continuum model fitting (\S\ref{subsec:uv}).

We have also analyzed a far-UV \emph{HST}/COS spectrum of NGC~7552 and measured properties of the outflow via absorption profiles.
We converted absorption profiles of \ion{Si}{ii} and \ion{Si}{iv} to a total hydrogen column density and recovered a total gas mass by assuming a simple radial shell geometry.
This yields a total mass contained within the wind of $M_{wind} = 2.0 \times 10^8$~M$_\odot$ with a mass outflow rate of $\dot{M}_{wind} = 19$~M$_\odot$~yr$^{-1}$, implying a mass loading factor of $\eta = 1.3$ -- 1.9.
We use a conservative estimate of the column density for this calculation; the true column --- and, similarly, its associated mass outflow rate --- could be 2 to 6 times higher.
This mass outflow rate is a factor of $\sim2$ higher than our mass outflow rate measured with the optical emission lines, once that value is corrected for extinction.

Compared directly, these two different analyses show that the UV and \Halpha\ emission have very different kinematic signatures and trace different regimes of the outflow.
The UV absorption profiles extend to velocities upward of $-1000$~\kms, far higher than seen in the \Halpha\ profile.
Our most liberal calculation of the maximum \Halpha\ outflow velocity has an average value of only $\langle\vmaxtwo\rangle = -290 \pm 100$~\kms.
As seen in Figure~\ref{fig:uv_vs_ha}, the majority of the gas column traced by the UV absorption exists at velocities larger than this.
There exists a large amount of material at high velocities that our \Halpha\ observations simply do not detect.
The ionization state of the gas inferred from the \ion{Si}{ii}/\ion{Si}{iv} ratio suggests that $> 99$ percent of the hydrogen is ionized.
The lack of \Halpha\ emission at these velocities is therefore likely due to a large drop in the local gas density, rather than a dearth of ionized hydrogen.
In short, \Halpha\ appears to be insensitive to a large fraction of the outflow mass of a similar gas phase.

It is somewhat surprising that we see strong, redshifted \Lyalpha\ emission along our line-of-sight (Fig.~\ref{fig:absorption}, lower-left).
In a galactic wind scenario, redshifted emission would be the result of backscattering from outflowing material on the far side of the galaxy.
But the disk, with an average extinction of $A_V = 5.3$, attenuates light by more than factor of a million at 1216~\AA, and thus no \Lyalpha\ emission from the far side of the disk would be present in our observations.
One logical explanation of this emission is that it traces material outflowing at a star-cluster scale, rather than tracing a bulk, galactic-scale outflow.
\citet{Leitherer13} model the \Lyalpha\ profile as a spherically-expanding shell with a constant expansion velocity.
They find a reasonable match to the emission profile using an expansion velocity of 50~\kms.
The model fails to account for substantial absorption at high velocities, however (see their figure~18).
One concern is that this excess absorption is the result of Galactic \Lyalpha\ absorption contaminating the line profile.
However, even the metal lines are nearly saturated beyond $-600$~\kms, where their model accounts for only $\sim50$ percent of the \Lyalpha\ absorption.
Therefore, the shell model and Galactic \Lyalpha\ contamination must not account for all the absorption we see at high velocities.
It is plausible, then, that the \Lyalpha\ simultaneously traces multiple outflow components: a low-velocity ($\sim 50$~\kms) cluster-scale outflow and a higher-velocity (up to 1000~\kms) galactic-scale wind.

Our interpretation of the \Halpha\ emission is that it likely arises on the cluster scale, similar to the \Lyalpha\ emission, rather than being associated with any global galactic wind.
In addition to the large disparities in the velocity profiles of \Halpha\ and various UV tracers, this interpretation is supported by the high electron densities seen in the broad emission lines.
Most of the broad \Halpha\ emission within the starburst ring comes from gas with electron densities of 300 -- 700~cm$^{-3}$, factors of a few higher than typically found in \ion{H}{ii} regions but consistent with other starbursts.
It is unlikely that any gas clumps entrained in a wind would maintain these high densities to large radii.
\citet{Werk14} infer typical gas densities for cool ($T = 10^4$~K) CGM of $n_e = 10^{-3}$ -- $10^{-4}$~cm$^{-3}$ at distances of 20 -- 150~kpc from the host galaxy.
It is more likely that these gas clumps exist near the disk and are being swept up in shocked supernovae outflows.
Based on the common picture of galactic winds --- hot gas rushing at high velocity past dense, warm clouds of gas --- the optical emission lines are likely arising from warm clouds that have not yet been accelerated by the wind.

Looking at the \Halpha\ emission in isolation, these clouds could either be material localized to star-forming regions and not yet entrained in the wind, or newly-entrained material at the base of the wind that has not yet been disrupted and accelerated to high velocities.
The broadening of the emission lines is then due to either turbulent mixing on the surface of the clumps \citep{Begelman90,Slavin93,Tenorio-Tagle10} or due to shock processes at the boundary of a local supernova bubble.
The line ratio diagnostics (Figs.~\ref{fig:voff}(e),(f)) demonstrate that the broad lines suggest shock-like ionization over much of the starburst ring.
The \SII/\Halpha\ line ratio, a more reliable single-line diagnostic, suggests that shocks could play a role in the gas ionization, with ratios near the shock ionization threshold ($\log($\SII/\Halpha$) \approx -0.5$) in all but the strongest regions of \Halpha\ flux.
Shock signatures can be diluted by strong photoionization in ground-based observations \citep{Hong13}, and our line ratio measurements are therefore likely to be underestimates.

Although we do not have information about how high above the disk this broad-line gas is located, we look at the nearby edge-on starburst galaxy M82 to shed light on how the \Halpha-emitting gas is distributed as a function of disk height in a starburst galaxy.
In Figure~\ref{fig:m82} we show the \Halpha\ profile of M82 as a function of height above the mid-plane, summed in a 1~kpc-wide box centered about the nucleus.
We use a narrow-band \Halpha\ image of M82 acquired with the Mini-Mosaic Imager on the WIYN 3.5~m telescope \citep{WestmoquetteThesis}.
We sum only the flux in the southern outflow, as the gas near the mid-plane in the northern outflow is partially obscured by the disk.
This narrow-band imagery does not distinguish between the narrow or broad components, but we find that the broad emission in NGC~7552 typically accounts for 50 percent of the total \Halpha\ flux.
If we assume this is also the case in M82, we can attribute half of the flux to the narrow emission.
If we further assume that the narrow emission arises in the disk, and thus all emission at low disk height is in the narrow component, then all narrow emission (50 percent of the total) is located within 160~pc of the mid-plane in M82.
Gas within 1~kpc accounts for 94 percent of the total \Halpha\ emission, or 87 percent of the remaining emission after discounting the first 50 percent attributed to the narrow component.
This supports the idea that a majority of the gas in the broad velocity component is located near the disk in starburst outflows.
This is also consistent with the high densities in the broad component, as it is unlikely that the $T = 10^4$~K gas would remain at such high densities out to large disk heights, and we therefore expect to preferentially detect gas near the disk.

The velocity centers of the broad component also support the idea that this component exists near the disk.
Comparing the velocity centers between the broad and narrow components in Fig.~\ref{fig:voff}(a) shows that the broad components still largely trace the rotation curve of the galaxy, despite their modest blueshifts.
Angular momentum conservation would be expected to substantially lower the rotational velocity of any outflowing gas as it reaches large radii.
The fact that the broad component still shows strong signs of rotation therefore suggests that the emitting gas remains close to the disk.

We have additionally created a toy model of an expanding spherical shell of gas, and compare the resulting velocity profile to our typical broad-component \Halpha\ line profile.
Coded in IDL, the model creates a three-dimensional cube of cells, with cells falling within defined inner and outer radii considered to have an arbitrary unit of gas in them.
We compute the column density for each cell by summing all the gas from that cell toward the observer.
We then normalize the column through the center of the sphere to the measured UV reddening $E(B - V) = 0.64$, and calculate the extinction toward each cell using a \citet{Cardelli89} model.
Each cell has a constant radial velocity that is projected into the line-of-sight, and then we create a velocity profile by constructing a histogram of the projected velocities.
We convolve this profile with a Gaussian profile representing the combined effects of our instrumental resolution ($\sigma = 17$~\kms) and turbulent/shock broadening.
We find that a constant-velocity shell with an expansion velocity of $100$~\kms\ that is turbulently-/shock-broadened by $\sigma = 90$~\kms\ creates a line-of-sight Gaussian velocity profile with $\mathbf{\sigma = 102}$~\kms\ ($\sim 240$~\kms\ FWHM) and a blueshifted centroid of $\mathbf{\Delta v = -48}$~\kms.
These values agree remarkably with our broad component measurements.

This model shows that the velocity center depends strongly on $E(B - V)$, and that the line width depends strongly on local broadening effects.
We have experimented with somewhat more-complex models by varying the radial density and velocity structures and find broadly similar results.
Varying the outflow velocity in the model leads to roughly linear changes in the velocity centroid offset, and an outflow velocity of $200$~\kms\ has a profile blueshifted by over $100$~\kms, much larger than what is seen in our data.
We also experimented with adding an additional velocity component by including a second shell with a different constant outflow velocity.
A second shell with a lower outflow velocity in our model creates a combined profile with a lower blueshifted centroid velocity, but that is still largely consistent with our data.
If the additional shell has a larger outflow velocity, the velocity centroid increases rapidly in response: a second shell with $v_\mathrm{out} = 200$~\kms\ creates a combined profile with a centroid of $-87$~\kms, inconsistent with our data.
Our model therefore suggests the line profile largely reflects the fastest outflow component, ultimately resulting in an over-estimate of the average outflow velocity of the gas. 

While this model may not be a perfect representation of the data, it demonstrates that simple scenarios involving low-velocity ($v \sim 100$~\kms) expanding gas shells can plausibly recreate the observed broad-component \Halpha\ kinematics with contributions from local broadening processes.
This model may not represent the exact physical scenario present in the star-forming region of NGC~7552, but we believe it demonstrates that such low-velocity scenarios need to be considered before adopting outflow velocities factors of a few larger than this model value.
If we adopt this model expansion velocity for the outflow velocity $v_\mathrm{out}$ in \S\ref{subsec:mass_outflow}, our calculated $\dot{M}_\mathrm{out}$ is reduced by nearly a factor of 3 to $\dot{M}_\mathrm{out} = 2.2$~M$_\odot$~yr$^{-1}$ ($\eta =$ 0.1 -- 0.2), a factor of 8.5 smaller than the UV measurement.
We believe this $v_\mathrm{out}$ represents a more accurate measurement of the bulk velocity of the gas, and therefore we adopt this adjusted $\dot{M}_\mathrm{out}$ as our best estimate of the mass outflow rate.

If the line widths of the broad \Halpha\ velocity component are being driven largely by surface layer turbulence or shock processes, then the line profile would contain little useful information about bulk motions of the gas.
Other recent observational work has reached similar conclusions about the mechanism behind the broad line widths.
\citet{Ho14} present an integral field study of a single isolated disk galaxy with similar global properties to NGC~7552.
The properties of the broad \Halpha\ components in their study largely resemble those which we find in NGC~7552, e.g., large line widths, high electron densities, and shocked line diagnostics.
They reach the conclusion that the line ratios and velocity dispersions of the broad velocity component are well-represented by shock models, and that compression along shock fronts leads to the high electron densities as measured via the \SII\ line ratio.

\begin{figure}
    \centering
    \includegraphics[width=\columnwidth]{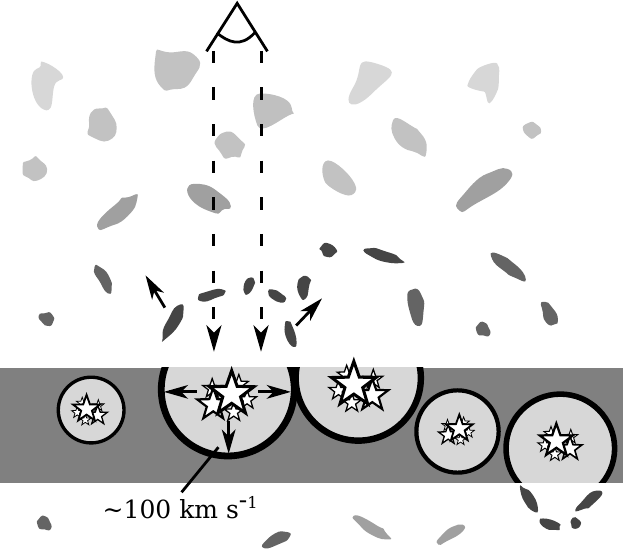}
    \caption{
        Schematic diagram of our interpretation of the broad nebular emission in NGC~7552.
        Star formation in closely-spaced clusters in the disk create hot, pressurized bubbles due to strong winds and supernovae.
        These bubbles can break out of the disk, launching dense material outward.
        These dense gas clumps expand as they travel to larger distances above the disk, lowering the local density and quickly diminishing the amount of \Halpha\ being emitted.
    The line widths of the nebular lines are likely the result of unresolved shocks or turbulent motions within these clumps, rather than representing the projection of the bulk flow velocity of the clumps.}
        \label{fig:schematic}
\end{figure}

We show a cartoon schematic of our physical picture in Figure~\ref{fig:schematic}.
Vigorous star formation is on-going in closely-spaced clusters in the disk, and strong stellar winds and supernovae create hot, pressurized bubbles within the disk, expanding at $\sim 50 - 100$~\kms.
These bubbles can break out of the top and bottom of the disk, launching dense material outward.
This gas is likely shocked, which drives the nebular line ratios and compresses the gas, leading to higher local gas densities compared to the star forming regions from which they originate.
These dense gas clumps are likely at high pressures ($n_e \sim 500$~cm$^{-3}$ and $T \sim 10^4$~K implies pressures of $P/k \gtrsim 10^6$), and would then expand as they travel to larger distances above the disk, lowering the local density and quickly diminishing the amount of \Halpha\ being emitted.

Recent simulations of galactic winds have also studied the warm gas in supernova-driven winds.
\citet{Creasey13} model a simplified ISM with star formation following the Kennicutt-Schmidt relation, with frequent supernova events in the disk driving gas into the halo.
They find that nearly all of the warm gas ($T = 10^4$~K) in their simulation exists in dense clumps at low velocities ($v \lesssim 350$~\kms).
They go on to note that the interface between the warm gas and the hot wind in their simulation could be the source of lower ionization lines like \SII.
\citet{Cooper09} model the evolution of dense, radiatively-cooling gas clouds within a hot wind.
Their results show such a cloud breaks into numerous small, dense ``cloudlets'' due to Kelvin-Helmholtz instabilities at the boundary between the dense gas and the hot wind.
These cloudlets typically exist at outflow velocities of $v \sim 200$~\kms\ in their simulation.
However, their emission-weighted $z$-velocity profile of the \Halpha\ line along a filament shows the emission peaking at 15~\kms.
This may be qualitatively similar to the picture painted by other \Halpha\ studies: the ``outflow velocity'' of the \Halpha\ gas is substantially larger than the velocity of the \Halpha\ profile centroid.
Neither study shows the presence of warm gas at high velocities, however, as seen in our UV data.
But neither study follows the gas out to large radii.

Overall, our results simply suggest that caution must be exercised when interpreting broad \Halpha\ velocity components as large-scale galactic winds.
The broad \Halpha\ component in NGC~7552 likely arises from dense material either being shocked or undergoing turbulent mixing.
It is still plausible --- even likely --- that the broad velocity component is tracing outflowing material which, combined with extinction, gives rise to the consistent $50$~\kms\ blueshift with respect to the disk component.
But our measurements suggest that the outflowing gas clumps traced by \Halpha\ in NGC~7552 have not yet been entrained in the larger wind.
They may be at the base of the wind, not yet having been accelerated to typical outflow velocities, or may be cluster-scale outflows still embedded in the disk.
In either case, it appears difficult to recover the true bulk flow rate from the observed \Halpha\ emission profile in the presence of turbulent motions and extinction.
Furthermore, the kinematics of the \Halpha\ profile are not representative of the high velocities that gas entrained in the galactic wind eventually reach.
As a result, UV absorption lines seem to more-completely trace the warm ionized galactic wind of NGC~7552.

\section{Conclusion}
\label{sec:conclusion}
We have analyzed spatially-resolved, medium-resolution, optical spectra of the barred spiral galaxy NGC~7552, using the emission lines \NII\ $\lambda\lambda$6548,6583; \Halpha; and \SII\ $\lambda\lambda$6716,6731.
We have also analyzed \emph{HST}/COS far-UV spectra along a sight-line coincident with our visible observations.
We summarize our main results as follows:

\begin{enumerate}
    \item We detect two velocity components in the optical spectra.
        We measure a narrow emission component (FWHM $\sim 100$~\kms) that is consistent with \ion{H}{ii} regions in other LIRGs.
        We detect an additional broad component that is typically blueshifted by $|\Delta v| =$ 30 -- 50~\kms and shows line widths of 240 -- 350~\kms\ FWHM.
        Both components exhibit high electron densities ($n_e \gtrsim 500$~cm$^{-3}$).

    \item Using methodology similar to other studies, we use the broad emission component to measure a mass outflow rate in \Halpha\ of $\dot{M}_\mathrm{out} = 5.0^{+2.7}_{-0.5}$~M$_\odot$~yr$^{-1}$.
        This implies a mass loading factor of $\eta =$ 0.3 -- 0.5.
        These values have been corrected for an extinction of $E(B - V) = 0.64$ measured from UV stellar continuum fitting and assuming a \citet{Calzetti00} extinction law.
        We caution, however, that we do not believe the kinematics of the \Halpha\ emission in NGC~7552 are consistent with large-scale mass outflow.
        A better estimate of the outflow velocity yields $\dot{M}_\mathrm{out} = 2.2$~M$_\odot$~yr$^{-1}$ ($\eta =$ 0.1 -- 0.2).

    \item Line ratio diagnostics suggest that the \Halpha-emitting gas could be at least partially shock-ionized ($\log($\SII/\Halpha$)$ $\approx -0.5$).
    Such shocks may be driving the large line widths of the broad velocity component.
        Turbulent broadening may also contribute to the large line widths.
        These additional broadening terms suggest that the line width is not representative of the outflow velocity (i.e., $v_\mathrm{out} \neq \Delta v_\mathrm{max}$).

    \item We independently measure the mass outflow rate using UV silicon absorption lines and find a rate of $\dot{M}_\mathrm{wind} = 19$~M$_\odot$~yr$^{-1}$ and an implied mass loading factor of $\eta = 1.3$ -- 1.9.
        This value is a factor of 8.5 higher than our best estimate in \Halpha.
        Our measurements may underestimate the gas column density N(H), which would lead to an underestimate of the UV-measured $\dot{M}_\mathrm{wind}$ by factors of a few.
    \item Although the \Lyalpha\ profile is consistent with a 50~\kms\ cluster-scale spherical outflow, such an outflow does not account for a large amount of high-velocity absorption.
        This high-velocity absorption is consistent with being material entrained in a galactic wind.
\end{enumerate}

Given the vigorous star formation on-going in NGC~7552 ($\dot{M}_\mathrm{SFR} \gtrsim 15$~M$_\odot$~yr$^{-1}$), the galaxy is likely producing a supernova-driven galactic wind.
It is not clear, however, that the \Halpha\ emission is an adequate tracer of this wind.
The optical emission lines show evidence of being shocked, suggesting that the broad linewidths are not the result of bulk gas flows, and outflow velocity estimates such as \vmaxtwo\ may be substantial over-estimates, leading to over-estimates of the mass outflow rate.
The high electron densities of this broad component also suggest it exists at small radius/disk height ($< 1$~kpc), rather than at $\sim5$~kpc as suggested by some authors.
A large column of UV-absorbing gas exists at high outflow velocities ($\Delta v \gtrsim 400$~\kms) that is undetected in our visible spectra, suggesting that the \Halpha\ is tracing material that is either in the inner regions of the galactic wind or is tracing local cluster outflows that may never escape the disk.
The \Lyalpha\ emission profile appears to trace similar regions, but strong absorption at high blueshifted velocities of both \Lyalpha\ and of low-ionization metal species shows that UV absorption is a more sensitive tracer of galactic outflows.
Accurately measuring the \Halpha\ outflow mass using this methodology likely requires substantial corrections to account for this undetected high-velocity gas.

\section*{Acknowledgments}
We thank the anonymous referee for graciously donating the time and effort to improving this manuscript.
This work was supported by NASA Headquarters under the NASA Earth and Space Science Fellowship Program - Grant NNX13AM34H, as well as by the National Science Foundation under Grant No.\ 0907839.
Some of the data presented in this paper were obtained from the Mikulski Archive for Space Telescopes (MAST).
STScI is operated by the Association of Universities for Research in Astronomy, Inc., under NASA contract NAS5-26555.
The authors also wish to acknowledge M.~Westmoquette for providing the imagery of M82.

\bibliography{ms.ngc7552}
\label{lastpage}
\end{document}